\documentclass[11pt]{article}
\usepackage[dvips]{epsfig}
\usepackage[T1]{fontenc}
\usepackage[latin1]{inputenc}
\usepackage{graphicx}
\usepackage[english]{babel}
\usepackage{amsmath}
\usepackage{amssymb}
\usepackage{amsfonts}
\usepackage[T1]{fontenc}
\usepackage{color}
\usepackage{babel}
\usepackage{verbatim}
\usepackage[unicode=true,pdfusetitle,bookmarks=true,bookmarksnumbered=false,bookmarksopen=false,
 breaklinks=false,pdfborder={0 0 1},backref=false,colorlinks=true]{hyperref}
\hypersetup{linkcolor=blue,citecolor=blue}
\makeatletter
\usepackage[dvips]{epsfig}
\usepackage[T1]{fontenc}
\usepackage{color}
\usepackage{pdflscape}
\usepackage{cite}

\title{\bf Quantum Cosmology in $f(R, T)$ Theory \\ with Schutz's Perfect Fluid}

\author{Serkan Doruk Hazinedar$^a$\thanks{
email: doruk.hazinedar@bilkent.edu.tr }, ~Yaghoub Heydarzade$^a$\thanks{
email: yheydarzade@bilkent.edu.tr}, ~and~ Shahram Jalalzadeh$^b$\thanks{
email: shahramjalalzadeh@iyte.edu.tr}\\
\\{\small $^a$Department of Mathematics, Faculty of Sciences, Bilkent University, 06800 Ankara, T{\" u}rkiye}
\\{\small $^b$Department of Physics, Izmir Institute of Technology, 35430, Izmir, T{\" u}rkiye}}
\date{February 13, 2026}

\begin{document}

\maketitle

\begin{abstract}
The $f(R, T)$ theory of gravity extends general relativity (GR) by allowing the gravitational Lagrangian to depend on both the Ricci scalar $R$ and the trace of the energy-momentum tensor $T$. The resulting matter-geometry coupling introduces additional dynamical effects that may account for the late-time acceleration of the universe without invoking dark energy. In the present work, we focus instead on the early-time regime and investigate the corresponding quantum cosmological dynamics. We analyze a Friedmann--Lemaitre--Robertson--Walker (FLRW) universe within the $f(R, T)$ framework, employing Schutz's perfect fluid formalism to extract a time parameter emerging from the matter sector itself. This approach is particularly well motivated in $f(R, T)$ gravity, where the coupling between geometry and the energy-momentum tensor's trace makes matter an active participant in the dynamics of spacetime and the evolution of cosmic time. The gravitational Hamiltonian, canonical momenta, and potential are derived, leading to the corresponding Schr\"odinger--Wheeler--DeWitt (SWDW) equation. The wave function of the universe is obtained for specific forms of $f(R, T)$, and the results are compared with previous studies in $f(R)$ and $f(R, T)$ models, highlighting the role of matter-geometry coupling in the emergence of quantum cosmological dynamics.
\end{abstract}

\maketitle
\vspace{0.5cm}

\section{Introduction}

 Understanding the gravitational interaction in regimes where quantum effects become significant remains one of the central open problems in theoretical physics. Although GR has achieved remarkable empirical success, it is expected to lose validity at sufficiently high curvature or energy scales, such as those characterizing the very early Universe, and thus calls for modification and/or quantization. Canonical quantum cosmology, based on the Wheeler--DeWitt (WDW) equation \cite{DeWitt1967, Wheeler1968}, offers a concrete setting in which such questions can be explored by quantizing the minisuperspace of homogeneous cosmological models. A central difficulty in this approach is the absence of an external time parameter (the ``problem of time''), which can be alleviated by introducing matter in the form of a perfect fluid. In particular, the canonical formalism developed by Schutz \cite{Schutz1970, Schutz1971} provides fluid variables that naturally yield an emergent internal time, thereby allowing the WDW equation to be cast into a Schr\"odinger-like form and enabling a probabilistic interpretation of the wave function of the universe \cite{Jalalzadeh:2020bqu, Alvarenga2002}.
 
On the other hand, cosmological observations, most notably the late-time accelerated expansion, have also motivated extensions of GR. A notable example is $f(R,T)$ gravity, proposed by Harko \textit{et al.}, in which the gravitational action is generalized to an arbitrary function of the Ricci scalar, $R$, and the trace of the energy-momentum tensor $T$ \cite{Harko2011}. The explicit dependence on $T$ introduces a nonminimal matter-geometry coupling, modifying both the gravitational field equations and, in general, the conservation properties of the matter sector, thereby enriching cosmological dynamics and phenomenology. Such models have been investigated extensively at the classical level, including reconstruction approaches \cite{ShabaniFarhoudi2013n}, stability analyses \cite{Alvarenga2013}, and viability studies \cite{Houndjo2012}.

From a more fundamental perspective, modified gravity is also well motivated in the high-curvature regime, where GR is expected to receive quantum corrections; higher-curvature terms can arise in semiclassical gravity and may affect the earliest stages of cosmic evolution. A classic illustration is the $R^2$ correction underlying Starobinsky inflation \cite{Starobinsky1980}. More broadly, $f(R)$-type extensions provide effective descriptions of departures from GR and have been reviewed extensively \cite{DeFeliceTsujikawa2010, SotiriouFaraoni2010}. These considerations make quantum cosmology in extended theories particularly compelling: minisuperspace quantization offers a controlled framework to examine how additional degrees of freedom and modified constraints may influence singularity behavior, semiclassical emergence, and the choice of boundary conditions for the universe. Along these lines, canonical and symmetry-based quantization schemes have been developed for several modified models, including $f(R)$ minisuperspace quantization via Noether symmetry and WDW methods \cite{Vakili2008PRD, CapozzielloLambiase2000}, as well as SWDW constructions in $f(R)$ cosmology employing Schutz's fluid variables to supply an internal time parameter \cite{VakiliQuadratic2009, VakiliScalarMetric2010}. Related minisuperspace quantizations have also been explored in other frameworks, such as Ho\v{r}ava--Lifshitz cosmology and teleparallel extensions \cite{PitelliLetelier2012, Darabi2019}. Within this broader program, the $f(R, T)$ setting is especially appealing because the matter sector enters explicitly in the gravitational Lagrangian, making the use of matter degrees of freedom as an emergent clock particularly well aligned with the structure of the theory; seminal WDW treatments of $f(R, T)$ quantum cosmology were presented in \cite{Harko qc}.

Motivated by these developments, in this work, we investigate FLRW quantum cosmology in $f(R, T)$ gravity using Schutz's canonical formulation of a perfect fluid as the matter source so that an internal time parameter arises directly from the matter sector. Our analysis focuses in particular on the minimally coupled class $f(R, T)=F^{0}(R)+G^{0}(T)$, for which the trace dependence is present but the mixed curvature-matter derivative vanishes, i.e., $f_{RT}=0$. This choice is physically well motivated and, at the same time, exposes an important structural distinction at the quantum level. In Ref.~\cite{Harko qc} a time parameter is introduced via $\tau=\int N(t)\,dt$, and a SWDW equation is obtained by identifying a specific Hamiltonian contribution associated with the $(D, M)$ sector (with $D\equiv f_{RT}$ and $M\equiv T-4p$, where $p$ is the thermodynamic pressure) as the momentum conjugate to $\tau$, a construction naturally adapted to models with nonvanishing $f_{RT}$ (e.g. $RT$-mixed couplings). On the other hand, our construction employs Schutz's fluid degrees of freedom to define time in a manner that remains directly applicable to the minimal $F^{0}(R)+G^{0}(T)$ sector. As a consequence, the SWDW evolution is obtained in a transparent canonical setting, with a well-defined inner product and Hermitian factor ordering. We then analyze the early universe limit and derive tractable reduced Hamiltonian and wave equations, obtaining explicit quantum states for representative choices of $F^{0}(R)$ and $G^{0}(T)$. This allows us to assess the imprint of matter-geometry coupling on minisuperspace dynamics, boundary-condition requirements (e.g., DeWitt-type suppression as $a\to0$), and the emergence of non-singular quantum behavior through wave-packet constructions.

The paper is organized as follows. In Sec.~2 we briefly review $f(R,T)$ gravity and formulate the FLRW minisuperspace dynamics, then introduce Schutz's perfect-fluid representation and derive the total Hamiltonian and its Dirac quantization, including a Hermitian factor-ordering prescription. In Sec.~3 we specialize to the minimal model $f(R,T)=F^{0}(R)+G^{0}(T)$ and develop the corresponding early-universe limit, obtaining the WDW and SWDW equations with the Schutz time variable. In Sec.~4 we study several representative choices of $F^{0}(R)$ and $G^{0}(T)$, construct explicit eigenstates and wave packets, and discuss the physical interpretation of the resulting quantum states. Appendix~A provides supporting analytic details on the early-universe behavior of the trace sector and the asymptotics used in the main text.


\section{$f(R, T)$ Theory and Canonical Quantization}
In this section, we shall consider an FLRW cosmology within the framework of $f(R, T)$ gravity. In the celebrated paper \cite{Harko2011}, the action for the $f(R, T)$ theory\footnote{We work in geometrized units in which 16$\pi$G=1=c.} is introduced as
\begin{equation}\label{1}
    S=\int_M f(R, T) \sqrt{-g} \mathrm{d}^4x + \int_M \mathcal{L}_{\text{matt}} \sqrt{-g} \mathrm{d}^4x,\
\end{equation}
where
$R$ is the scalar curvature, $T$ is the trace of the energy-momentum tensor, $f(R, T)$ is an arbitrary function of
$R$ and $T$, and $\mathcal{L}_{\text{matt}}$ is the matter Lagrangian density of the theory. It is also assumed that $\mathcal{L}_{\text{matt}}$ does not depend on the derivative of the metric. 

Then, the field equations of $f(R, T)$ gravity theory read \cite{Harko2011}
\begin{eqnarray}\label{field eq 1}
    f_R R_{\mu\nu} -\frac{1}{2} f g_{\mu\nu} +(g_{\mu\nu}\Box -\nabla_{\mu}\nabla_{\nu}) f_R= \frac{1}{2} T_{\mu\nu} -f_T  T_{\mu\nu} -f_T  \Theta _{\mu\nu},
\end{eqnarray}
where $f\equiv f(R, T)$, $f_R= \partial_R f$, $f_T= \partial_T f$, and $\Theta_{\mu\nu} \equiv g^{\alpha\beta} \delta T_{\alpha\beta}/ \delta g^{\mu\nu}$. Contracting the above field equations leads to the following equation 
\begin{eqnarray}
\label{field eq 2}
f_R R +3~\Box f_R  -2f=
\frac{1}{2} T -f_T T -f \Theta .    
\end{eqnarray}

For a perfect fluid, in the comoving frame,  the energy-momentum tensor takes the following form,
\begin{equation}
T_{\mu\nu}=(\rho + p)U_{\mu}U_{\nu}+pg_{\mu\nu},
\end{equation}
where $\rho$ and $p$ are the matter energy density and thermodynamic pressure, respectively, and $U^{\mu}= \delta_0^{\mu}$ is the 4-velocity of the comoving observer, which satisfies the timelike normalization condition $U^{\mu}U_{\mu} =-1$. 
 Moreover, in Schutz's formalism, the matter Lagrangian density is fixed as $\mathcal{L_{\text{matt}}}=p$ for the perfect fluid matter model, and it yields $\Theta_{\mu\nu} =-2T_{\mu\nu}+pg_{\mu\nu}$. \cite{Schutz1970, Schutz1971, Harko2011, Pedram, Harko qc} Finally, the theory has an {energy-balance equation} \cite{Harko2011, Harko qc}
\begin{equation}\label{energy-balance}
    U_{\mu}\nabla^{\mu}\rho +(\rho +p)\nabla ^{\mu}U_{\mu} =-\frac{f_T}{\frac{1}{2} +f_T }
 \Bigg[(\rho +p)U_{\mu} \nabla^{\mu}\ln f_T +\frac{1}{2}U_{\mu}\nabla^{\mu}(\rho -p)\Bigg],
\end{equation}
which shows that fluid energy-momentum is not covariantly conserved in $f(R,T)$ gravity due to
$f_{T}$-dependent source terms. Note that using $\mathcal{L}_{\mathrm{matt}}=-\rho$ modifies $\Theta_{\mu\nu}$ and the
non-conservation law affects the minisuperspace coupling to $T$. Since Schutz's Hamiltonian
formulation is naturally compatible with $\mathcal{L}_{\mathrm{matt}}=p$, we maintain this choice throughout the paper. See also \cite{boh}, where the gravitational field equations for theories with trace couplings to a relativistic perfect fluid were rederived using both the Brown and Schutz variational formalisms, and it was noted that certain results in the literature may benefit from further careful analysis.

In the present analysis, we adopt the space-time geometry described by the FLRW metric. In spherical coordinates, the line element reads
\begin{equation}\label{FRW}
ds^{2} = -N^{2}(t)\,dt^{2} + a^{2}(t)\!\left[\frac{dr^{2}}{1 - k r^{2}} + r^{2}\!\left(d\theta^{2} + \sin^{2}\!\theta\, d\varphi^{2}\right)\right],
\end{equation}
where $N(t)$ is the lapse function, $a(t)$ denotes the cosmological scale factor, and $k \in \{1,0,-1\}$ specifies the spatial curvature, corresponding to closed, flat, and open Universe models, respectively.
Following the notation in \cite{Harko qc}, we define 
\begin{equation}\label{variables}
\begin{array}
{rcl rcl rcl rcl} A&:=&f_R, & B&:=&f_T, & C&:=&f_{RR}, & D&:=&f_{RT},\\ 
E&:=&f_{TT}, & F&:=&\Box f_{RR}, & G&:=&\Box f_{RT}, & M&:=&T-4p.
\end{array}
\end{equation}

To obtain the minisuperspace Hamiltonian, we substitute the FLRW metric \eqref{FRW}
into the action \eqref{1} and perform integration over the spatial coordinates. This
yields a reduced Lagrangian, $\mathcal{L}(a,\dot a,R,\dot R,T,\dot T)$,
whose explicit form is given in \cite{Harko qc}. Following the Palatini--Ostrogradsky procedure, we introduce the auxiliary variables $(A,B,C,D,E,F,G,M)$ defined in \eqref{variables} in order to linearize the higher-derivative
terms and write the kinetic sector in first-order canonical form. 
Performing a Legendre transformation
\[
H_{\rm grav}=\sum_I P_I \dot I - \mathcal{L},
\]
where $I \in\{a, A, B, C, D, E, F, G, M\}$ and eliminating the velocities $\dot I$, where $\dot{(I)} \equiv \mathrm{d}(I)/\mathrm{d}t$, in favour of the momenta then gives
the gravitational Hamiltonian quoted in \eqref{eq.8} as follows 
\begin{eqnarray}\label{eq.8}
H_{grav} = &+&\frac{N}{6a^3} \left(P_A +\frac{1}{3}P_F\right)(A P_A +F P_F -a P_a)\nonumber\\
&+&\frac{N}{6a^3} \frac{\mathcal{A}}{B} \left(P_C P_R + P_D P_M +\frac{P_E P_F}{3} E+\frac{P_G^2}{9}  RE\right)\nonumber\\
&+&\frac{N}{6a^3}2P_C \left(P_A +\frac{1}{3}P_F\right)C +Na^3 V-6kNa \left(A -\frac{B\mathcal{Z}}{\mathcal{A}}\right),
\end{eqnarray}
where the potential, $V$, is given by
\begin{equation}
V=-f +\widetilde{\lambda}R +\widetilde{\mu}\left(\frac{1}{2} T +B M -A R -3\Box A +2f \right),
\end{equation}
and
\begin{equation}\label{Hamiltonian analysis}
\begin{split}
&    \mathcal{A}= \tfrac12 + 3B + E M - D R - 3G,~~~~~~~~
    \mathcal{Z}=DM-CR+A-3F,\\
&    \widetilde{\mu}= \frac{\mu}{Na^3}=  \frac{B}{\mathcal{A}},~~~~~~~~
    \widetilde{\lambda}= \frac{\lambda}{Na^3}=A - \frac{B\mathcal{Z}}{\mathcal{A}}.
    \end{split}
\end{equation}
 Here, $\mu$ and $\lambda$ are Lagrange multipliers, and $P_I$ is the canonical momentum of $I$.

 The canonical
momenta $P_I$ associated with each variable $I$ are
computed via $P_I=\partial\mathcal{L}/\partial\dot I$, yielding 
\begin{equation}
\begin{split}\label{defA-F}
&P_{a} = -\frac{12}{N}\,a\,\dot a\,\widetilde{\lambda}-\;\frac{6}{N}\,a^{2}\,\dot{\widetilde{\lambda}},~~~
P_{A} = -\frac{6}{N}\,a^{2}\,\dot a \left(1-\frac{B}{\mathcal{A}}\right),\\
&P_{B} = -\frac{6}{N}\,a^{2}\,\dot a \left(-\frac{\mathcal{Z}}{\mathcal{A}}+\frac{3B\mathcal{Z}}{\mathcal{A}^{2}}\right),~~~
P_{C} = -\frac{6}{N}\,a^{2}\,\dot a \left(\frac{BR}{\mathcal{A}}\right),\\
&P_{D} = -\frac{6}{N}\,a^{2}\,\dot a \left[-\frac{B(T-4p)}{\mathcal{A}}-\frac{B\mathcal{Z}R}{\mathcal{A}^{2}}\right],~~~ 
P_{E} = -\frac{6}{N}\,a^{2}\,\dot a \left[\frac{B\mathcal{Z}(T-4p)}{\mathcal{A}^{2}}\right],
\\
&P_{F} = -\frac{6}{N}\,a^{2}\,\dot a \left(\frac{3B}{\mathcal{A}}\right),~~~
P_{G} = -\frac{6}{N}\,a^{2}\,\dot a \left(-\frac{3B\mathcal{Z}}{\mathcal{A}^{2}}\right),\\
&P_{R} = -\frac{6}{N}\,a^{2}\,\dot a \left[\frac{BC}{\mathcal{A}}-\frac{B\mathcal{Z}D}{\mathcal{A}^{2}}\right],~~~
P_{T} = -\frac{6}{N}\,a^{2}\,\dot a \left[-\frac{BD}{\mathcal{A}}+\frac{B\mathcal{Z}E}{\mathcal{A}^{2}}\right],\\
&P_{p} = -\frac{6}{N}\,a^{2}\,\dot a \left(\frac{4DB}{\mathcal{A}}-\frac{4B\mathcal{Z}E}{\mathcal{A}^{2}}\right).
\end{split}
\end{equation}
Note that the cross-terms proportional to $P_AP_F$ and $P_CP_R$, as well as the
coefficients $\mathcal{A}/B$ and $C,E,G$ in \eqref{eq.8}, arise from inverting the kinetic matrix
associated with the higher derivatives of $f(R,T)$, as discussed in
\cite{Harko qc}. We refer the reader to
\cite{Harko qc} for the detailed algebraic structure of the minisuperspace kinetic
matrix and for the consistency of this reduction.

\noindent In Schutz's formalism, the canonical 1-form contains $P_{\tau}\dot{\tau}$ and the reduced FLRW
matter Hamiltonian takes the form $H_{\mathrm{matt}}=Na^{-3\alpha}P_{\tau}$. Matching this to 
   $ H_{\text{matt}}= -\sqrt{-g}\mathcal{L}_{matt}= - Na^3 p$,
as in the celebrated works \cite{Schutz1970, Schutz1971,  Lapchinskii, Vakali}, yields the identification 
   $ -N a^3 p \to N a^{-3 \alpha}P_{\tau}$.
Equivalently, for non-vanishing $N$ and $a$, the transformation is reduced to
\begin{eqnarray}\label{transform}
    p \to -a^{-3(\alpha+1)}P_{\tau},
\end{eqnarray}
which is consistent with the Schutz barotrope. The Poisson brackets remain canonical, and the barotropic
equation of state is preserved in minisuperspace. Here, $\alpha$ is the equation of state parameter, where $\tau$ is a function of thermodynamical entropy and the $\epsilon$-potential in the Schutz' representation of perfect fluid described in \cite{ Schutz1971, Vakali}. Hence, the barotropic equation of state is modified according to the canonical transform in \eqref{transform} as follows
\begin{eqnarray}\label{equation of state}
    \alpha \rho = p \to -a^{-3(\alpha+1)}P_{\tau}.
\end{eqnarray}
Consequently, both the energy density and the trace of the energy-momentum tensor become
\begin{eqnarray}\label{rhoTrelation}
\rho &=& -\frac{1}{\alpha} a^{-3(1+\alpha)} P_{\tau},\\
T &=& -\rho + 3p = -\frac{1 - 3\alpha}{\alpha}\, a^{-3(1+\alpha)} P_{\tau},
\end{eqnarray}
when $\alpha \neq 0$. Since the geometric coefficients $A, B, C, D, E, F, G,$ and $M$ in the gravitational Hamiltonian are functions of $T$, this transformation implies that all these coefficients acquire an implicit dependence on both $a$ and $P_{\tau}$. Therefore, the matter-geometry coupling in $f(R,T)$ gravity becomes dynamically encoded in the canonical variables, and the trace $T$ acts as a mediator between the fluid momentum and the geometric sector.
Then, the total Hamiltonian, $H$, for $f(R, T)$ gravity theory in Schutz's representation reads as follows: 
\begin{eqnarray}\label{total Hamiltonian}
 H= H_{\text{grav}} + H_{\text{matt}}
 &=&\frac{N}{6a^3} \left(P_A +\frac{1}{3}P_F\right)(A P_A +F P_F -a P_a)\nonumber\\
&-&\frac{N}{6a^3} \frac{\mathcal{A}}{B} \left(P_C P_R + P_D P_M +\frac{P_E P_F}{3} E+\frac{P_G^2}{9}  RE\right)\nonumber\\
&+&\frac{N}{6a^3}2P_C \left(P_A +\frac{1}{3}P_F\right)C
+Na^3 V-6kNa \left(A -\frac{B\mathcal{Z}}{\mathcal{A}}\right)+ N a^{-3\alpha}P_{\tau}.
\end{eqnarray}

\noindent Classical dynamics is governed by the Hamilton equations. The time evolution of each parameter can be obtained as
\begin{eqnarray*}
    \dot{a}&=& \{a, H\}=- \frac{N}{6a^3}\left( P_A + \frac{1}{3} P_F \right)a,\nonumber\\
    \dot{A}&=& \{A, H\}= \frac{N}{6a^3}\Big[2AP_A+\left( F+ \frac{1}{3}A\right)P_F+ 2P_C C- aP_a \Big],\nonumber\\
    \dot{B}&=&\{B, H\}= 0,\nonumber\\
\end{eqnarray*}
\begin{eqnarray}\label{gauge}
 \dot{C}&=&\{C, H\} = -\frac{N}{6a^3}\Big[\frac{\mathcal{A}}{B}P_R -2 \left( P_A + \frac{1}{3} P_F \right) C\Big],\nonumber\\
    \dot{D}&=& \{D, H\} =-\frac{N}{6a^3}\left( \frac{\mathcal{A}}{B} P_M \right),\nonumber \\
    \dot{E}&=&\{E, H\} = -\frac{N}{6a^3}\left( \frac{\mathcal{A}}{B} \frac{1}{3}P_F E \right),\\
 \dot{D}&=& \{D, H\} =-\frac{N}{6a^3}\left( \frac{\mathcal{A}}{B} P_M \right), \nonumber \\
    \dot{E}&=&\{E, H\} = -\frac{N}{6a^3}\left( \frac{\mathcal{A}}{B} \frac{1}{3}P_F E \right),\nonumber \\
   \dot{F}&=& \{F, H \} =\frac{N}{18a^3}\Big[ AP_A + 2 FP_F -aP_a + 3P_A F + \frac{\mathcal{A}}{B}P_C C-\frac{\mathcal{A}}{B} P_E E\Big],\nonumber \\
    \dot{G}&=&\{G, H \} = -\frac{N}{6a^3}\left( \frac{\mathcal{A}}{B} \frac{2}{9} P_G R E \right),\nonumber \\
    \dot{M}&=& \{M, H \} =-\frac{N}{6a^3}\left( \frac{\mathcal{A}}{B} P_D \right),\nonumber \\
    \dot{\tau}&=&\{ \tau, H\} = \frac{N}{a^{3\alpha}},\nonumber 
\end{eqnarray}
where a dot denotes the time derivative. To determine the time at the classical level, we may remove the gauge freedom via the fixing of the gauge. Then, one can obtain the following from \eqref{gauge} 
\begin{equation}\label{tau}
    N= a^{3\alpha} \implies \tau =t,
\end{equation}
which means $\tau$ will play the role of time in the model. The Dirac quantization of the gravitational Hamiltonian of the $f(R, T)$ theory is obtained in \cite{Harko qc}, in which the ordering parameter method is used to obtain a Hermitian super-Hamiltonian. In the quantization of the minisuperspace Hamiltonian, one must specify an operator ordering between the non-commuting variables $q$ and their conjugate momenta $P_q$, since $[q, P_q] = i\hbar$. Following \cite{Harko qc, PedramJalalzadeh2008}, we adopt an ordering procedure that guarantees the Hermiticity of the Hamiltonian operator. In this framework, the ordering ambiguity is parameterized by three real numbers $(u,v,w)$ satisfying $u+v+w=1$, and the momentum operator $P_q = -i\partial/\partial q$ acts according to
\begin{equation}\label{orderingHarko1}
q P_q^{2}
=\tfrac{1}{2}\left(q^{u} P_q q^{v} P_q q^{w} + q^{w} P_q q^{v} P_q q^{u}\right)
= -q\frac{\partial^{2}}{\partial q^{2}} + \frac{u w}{q},
\end{equation}
where the parameters $u$ and $w$ encode the factor ordering ambiguity of the kinetic term.
Similarly, for the first-order product, one defines
\begin{equation}\label{orderingHarko2}
q P_q
=\tfrac{1}{2}\left(q^{r} P_q q^{s} + q^{s} P_q q^{r}\right)
= -i\left(q\frac{\partial}{\partial q} + \frac{1}{2}\right),
\qquad r+s=1,
\end{equation}
and for the second-order product one defines
\begin{eqnarray}
    q^{-2} P_q = -i \left( - \frac{2}{q^3}+ \frac{1}{q^2} \frac{\partial}{\partial q}\right) \qquad r+s =-2,
\end{eqnarray}
where $(r,s)$ parameterize the ordering of $q$ and $P_q$.
Then, combining the resulting equation with the quantized matter Hamiltonian in Schutz's representation leads
\begin{eqnarray}\label{constarint}
\mathcal{H} = &-&\frac{1}{6a^3}\Bigg[\left(A\frac{\partial}{\partial A} +F\frac{\partial}{\partial F} +2C\frac{\partial}{\partial C} -a\frac{\partial}{\partial a}+5\right)\Bigg(\frac{\partial}{\partial A}+\frac{1}{3}\frac{\partial}{\partial F}\Bigg) -u_1 w_1 \frac{1}{A}-u_2 w_2 \frac{1}{3F}\Bigg]\nonumber\\
&+& 
\frac{1}{6a^3}\Bigg[\frac{\mathcal{A}}{B}\Bigg( \frac{\partial}{\partial C}\frac{\partial}{\partial R}  +\frac{\partial}{\partial D}\frac{\partial}{\partial M}+ \frac{E}{3}\frac{\partial}{\partial E} \frac{\partial}{\partial F} - \frac{RE}{9}\frac{\partial ^2}{\partial G ^2} \Bigg)\nonumber\\
&+&  \Bigg(- \frac{D}{B}\frac{\partial}{\partial C} -\frac{R}{B}\frac{\partial}{\partial M} +\frac{E}{B}\frac{\partial}{\partial D} +
\frac{\mathcal{A}+EM}{3B}\frac{\partial}{\partial F}   +\frac{RE}{3BG}u_3w_3      \Bigg) \Bigg]\nonumber\\
&+& a^3 V -6ka\Bigg(A -\frac{B\mathcal{Z}}{\mathcal{A}}\Bigg) -ia^{3\alpha}\frac{\partial}{\partial \tau}.
\end{eqnarray}
Here, $u_1,w_1$, $u_2,w_2$ and $u_3,w_3$ denote the ambiguity in the ordering of factors $A , P_A$, $F, P_F$, and $G,P_G$, respectively. With the inner product $\langle\Phi|\Psi\rangle=\int a^{-3\alpha}\Phi^{*}\Psi\,da\,dA$, Hermiticity
requires $u_{i}w_{i}\ge0$, where $i\in \{1,2,3\}$, and vanishing boundary currents. We adopt $u_{i}=w_{i}\in[0,1]$,
ensuring symmetric kinetic operators and positive ordering shifts.

In the following, we present several remarks that clarify the quantization procedure and constraint analysis.

\noindent
\textbf{Remark 1}: According to the Dirac quantization method described in \cite{Dirac1958, Dirac1959, Dirac1964}, the super-Hamiltonian $\mathcal{H}$ in \eqref{constarint} must weakly vanish, $\mathcal{H} \approx 0$ and the primary constraint $H\approx0$ generates time reparameterizations. The gauge condition
$\chi\equiv N-a^{3\alpha}=0$ satisfies $\{\chi,H\}\neq0$ on the constraint surface, forming an
admissible second-class pair. Requiring $\dot{\chi}\approx0$ introduces no new constraints, closing the Dirac algebra consistently.

\noindent
\textbf{Remark 2:} From \eqref{tau}, one obtains \(\dot{\tau}=1>0\). Therefore, \(\tau\) is strictly monotonic for all trajectories with \(a>0\), including expanding, contracting, and bouncing solutions. Note that a bounce occurs at \(\dot a=0\), not at \(a=0\)). In the quantum theory, admissible wave packets are supported on the half-line \(a>0\), so the SWDW evolution in \(\tau\) is well defined globally.

\noindent
\textbf{Remark 3:} In FLRW cosmology, all fields depend only on the time coordinate. The variables \((A,B,C,\dots)\) are algebraically defined by \(f(R,T)\) and their canonical momenta \eqref{gauge} enforce precisely these algebraic relations. The time preservation of the primary constraints closes on the set already present; no new second-class constraints arise. Thus, the minisuperspace reduction is consistent and complete.

\section{The Quantum Cosmology of the ${f(R,T)=F^{0}(R)+G^{0}(T)}$ Model:}
\hfill

In this section, we shall consider the particular class of models characterized by a gravitational Lagrangian of the form
\begin{equation}\label{minimal}
f(R,T) = F^{0}(R) + G^{0}(T),
\end{equation}
where $F^{0}(R)$ and $G^{0}(T)$ are arbitrary, sufficiently smooth functions of the Ricci scalar $R$ and the trace of the energy-momentum tensor $T$, respectively. For this choice, the partial derivatives of the action with respect to $R$ and $T$ and their second-order counterparts are given by
\begin{equation}
\begin{array}{rcl rcl rcl rcl}
A &:=& F^{0}_{R}(R),&
B &:=& G^{0}_{T}(T),&
C &:=& F^{0}_{RR}(R),
\\[6pt]
D &:=& 0,&
E &:=& G^{0}_{TT}(T),&
F &:=& \Box F^{0}_{RR}(R),
\\[6pt]
G &:=& 0,&
M &:=& T + 4a^{-(3\alpha +1)}P_{\tau}.
\end{array}
\end{equation}
In terms of the above definitions, the function $\mathcal{A}$ introduced in the Hamiltonian analysis \eqref{Hamiltonian analysis} takes the form
\begin{align}
\mathcal{A}=\frac{1}{2} + 3B + E\,M - D\,R - 3G =\frac{1}{2} + 3\,G^{0}_{T}(T) + G^{0}_{TT}(T)\,\bigl(T - 4p\bigr).
\end{align}
Here, one notes that $D = 0$ and $G = 0$ due to the minimal coupling in the Lagrangian \eqref{minimal}.
Finally, the ratio $B/\mathcal{A}$, which enters the gravitational Hamiltonian \eqref{total Hamiltonian}, is expressed as
\begin{equation}
\frac{B}{\mathcal{A}}
= \frac{G^{0}_{T}(T)}{\frac{1}{2} + 3\,G^{0}_{T}(T) + G^{0}_{TT}(T)\,\bigl(T - 4p\bigr)} \, .
\end{equation}
The above relations demonstrate that the general minisuperspace Hamiltonian and its quantized version derived in the preceding section can be applied directly to this class of $f(R,T)$ models by substituting the specialized expressions of $A,B,C,D,E,F,G,$ and $M$. The novelty of this work lies in deriving the time variable \(\tau\) from the Schutz matter sector within \(f(R,T)\) gravity, 
allowing a consistent SWDW evolution not present in previous formulations. In the paper \cite{Harko qc}, it is assumed that the Ricci scalar diverges, $R \to \infty$, in the early universe. One can observe that the Ricci scalar, $R$, and the trace of the energy-momentum tensor, $T$, are implicitly dependent on each other from \eqref{field eq 2}. Therefore, in the early universe, the ratio $B/\mathcal{A}$ should converge to a scalar-valued function $\varphi$, that is, 
\begin{eqnarray}\label{converge}
    \frac{B}{\cal A} \to \varphi,
\end{eqnarray}
when $ R\to\infty$ and $\mathcal{A}\neq 0$. Hence, we have the following immediate remarks.

\noindent\textbf{Remark 4:} When $B/\mathcal{A}$ is a scalar, then $B/\mathcal{A}=\varphi$ in the limit of $R \to \infty$.

\noindent \textbf{Remark 5:} For typical choices \(F^0_R\sim R^{n-1}\) and \(G^0_T\sim |T|^{m-1}\) (with \(n\ge2\)), the traced field equations imply \(|T|\to\infty\) as \(R\to\infty\). Then
\[
\frac{B}{\mathcal{A}}=\frac{G^0_T(T)}{F^0_R(R)}\sim \frac{|T|^{m-1}}{R^{n-1}}\to\varphi\in\mathbb{R},
\]
depending on the dominant asymptotic behavior. The special examples in Section~4 yield \(\varphi=0\) or \(\varphi=2/7\).

\noindent From the definitions of $P_A$ and $P_F$ given in \eqref{defA-F}, one can obtain the following:  
\begin{eqnarray}
    P_F = \left(\frac{3B}{\mathcal{A}-B}\right)P_A.
\end{eqnarray} 
Consequently, in the early universe limit, the total Hamiltonian on the minisuperspace $(I, \tau)$, where $I \in \{a, A, B, C, D, E, F, G, M\}$, reads
\begin{eqnarray}\label{limit Hamiltonian}
H\rightarrow&+& \frac{N}{6a^3} P_A \Bigg(\frac{1}{1-\varphi}\Bigg) (A P_A +F P_F -a P_a)\nonumber\\
&-&
\frac{N}{6a^3} \frac{1}{\varphi} \Bigg(P_C P_R + P_D P_M +\frac{\varphi}{1-\varphi}P_E P_A E+\frac{P_G^2}{9}  RE\Bigg)\nonumber\\
&+& \frac{N}{6a^3}\left( \frac{1}{1-\varphi}\right)2P_CP_AC +Na^3 V-6kNa \Big[(1-\varphi)A-\varphi(CR+3F)\Big ]+Na^{-3\alpha}P_{\tau},
\end{eqnarray}
where the potential converges as follows
\begin{equation}
    V \to (AR-f) + \varphi \left(\frac{1}{2}T+BM+CR^2 -3FR-2AR+2f\right).
\end{equation}
Here, we used the Stokes theorem to vanish $\Box f_R$-term in the potential $V$
\begin{equation}\label{vanish}
    \int_M Na^3\Box f_R dt d^3x = \int_M \Box f_R\sqrt{-g}d^4x=\int_{\partial M} f_R^{;\mu} \sqrt{h}d^3 \sigma=0, 
\end{equation}
where $h$ is the determinant of the induced metric $h_{\mu \nu}$ and $d\sigma$ is the volume form of the hypersurface $\partial M$. Indeed, for compact spatial sections, such as \(\Sigma=S^3\), and fixed initial/final time slices, the boundary integral \(\int_{\partial M} f_R^{;\mu}n_\mu\sqrt{h}\,d^3\sigma\) vanishes. For noncompact \(\Sigma\), standard Dirichlet fall-off \(f_R^{;\mu}n_\mu\rightarrow 0\) ensures the same. Both are standard assumptions in minisuperspace quantum cosmology. Note that in quantum cosmology, we usually assume there is no boundary for the universe, which means $\Sigma$ is compact. Moreover, the following equation
\begin{equation}\label{difference}
    \frac{1}{3a^3} P_D P_M=h D P_D=0,
\end{equation}
is  obtained in \cite{Harko qc}. Here, $h= \frac{1}{N}\frac{\dot a}{a}$ is the Hubble parameter. Then, due to \eqref{difference}, the minisuperspace Hamiltonian in the early universe limit \eqref{limit Hamiltonian} reduces to
\begin{eqnarray}\label{reduced limit}
    H \to &+&\frac{N}{6a^3} P_A \left(\frac{1}{1-\varphi}\right) (A P_A +F P_F -a P_a) -\frac{N}{6a^3} \frac{1}{\varphi} \Bigg(P_C P_R +\frac{\varphi}{1-\varphi}P_E P_A E+\frac{P_G^2}{9}  RE\Bigg)\nonumber\\ 
&+& \frac{N}{6a^3}\left( \frac{1}{1-\varphi}\right)2P_CP_AC +Na^3(AR-f)+Na^3\varphi \Bigg(\frac{1}{2}T+BM+CR^2 -3FR-2AR+2f \Bigg)\nonumber\\
&-&6kNa \left((1-\varphi)A \right)
+18kNa\varphi F+6kNa\varphi CR+Na^{-3\alpha}P_{\tau}. 
\end{eqnarray}

Here, one should note the following remarks.

\noindent \textbf{Remark 6:} In \cite{Harko qc}, the effective time is defined by $\tau = \int_M N(t) dt$, or, equivalently, $d\tau/dt = N(t)$. It leads to performing the transformation $-\frac{1}{3a^3}P_D P_M \rightarrow P_{\tau}$. However, this transformation is not applicable to the minimal model \eqref{minimal} of $f(R, T)$  due to \eqref{difference}. This is a significant difference for the minimal and non-minimal models for $f(R, T)$ while considering the time problem.

\noindent Then, the corresponding super-Hamiltonian converges as follows: 
\begin{eqnarray}
\mathcal{H} &\to& \frac{1}{6 a^{3}(1-\varphi)}
\left(
a\,\frac{\partial^{2}}{\partial a\,\partial A}
- A\,\frac{\partial^{2}}{\partial A^{2}}
+ u_{1}w_{1}\,\frac{1}{A}
- F\,\frac{\partial}{\partial F}\frac{\partial}{\partial A}
- 2\,\frac{\partial}{\partial A}
\right) \nonumber\\
&& + \frac{1}{6 a^{3}\varphi}
\Big(
\frac{\partial}{\partial C}\frac{\partial}{\partial R}
+ R E\,\frac{\partial^{2}}{\partial G^{2}}
\Big) + \frac{1}{6 a^{3}(1-\varphi)}
\Big(
E\,\frac{\partial}{\partial E}\frac{\partial}{\partial A}
+ 2\,\frac{\partial}{\partial A}
\Big) \nonumber\\
&& - \frac{1}{3 a^{3}(1-\varphi)}
\Big(
C\,\frac{\partial}{\partial A}\frac{\partial}{\partial C}
+ 2\,\frac{\partial}{\partial A}
\Big) + a^{3}\,(A R - f) \nonumber\\
&& + a^{3}\varphi \bigg(\frac{1}{2}T + B M + C R^{2} - 3 F R - 2 A R + 2 f \bigg) \nonumber\\
&& - 6 k a \big((1-\varphi) A\big) + 18 k a\,\varphi F + 6 k a\,\varphi C R - i\,a^{-3\alpha}\frac{\partial}{\partial \tau}\, .
\end{eqnarray}
Therefore, the corresponding WDW equation reads
\begin{eqnarray}\label{eq.58}
\mathcal{H}\Psi=\Bigg\{&+& \frac{1}{6 a^{3}(1-\varphi)}
\Big(
a\,\frac{\partial^{2}}{\partial a\,\partial A}
- A\,\frac{\partial^{2}}{\partial A^{2}}
+ u_{1}w_{1}\,\frac{1}{A}
- F\,\frac{\partial}{\partial F}\frac{\partial}{\partial A}
- 2\,\frac{\partial}{\partial A}
\Big) \nonumber\\
&+& \frac{1}{6 a^{3}\varphi}
\Big(
\frac{\partial}{\partial C}\frac{\partial}{\partial R}
+ R E\,\frac{\partial^{2}}{\partial G^{2}}
\Big) + \frac{1}{6 a^{3}(1-\varphi)}
\Big(
E\,\frac{\partial}{\partial E}\frac{\partial}{\partial A}
+ 2\,\frac{\partial}{\partial A}
\Big) \nonumber\\
& -& \frac{1}{3 a^{3}(1-\varphi)}
\Big(
C\,\frac{\partial}{\partial A}\frac{\partial}{\partial C}
+ 2\,\frac{\partial}{\partial A}
\Big) + a^{3}\,(A R - f) \nonumber\\
& +& a^{3}\varphi \bigg(\frac{1}{2}T + B M + C R^{2} - 3 F R - 2 A R + 2 f \bigg) \nonumber\\
& -& 6 k a \big((1-\varphi) A\big) + 18 k a\,\varphi F + 6 k a\,\varphi C R - i\,a^{-3\alpha}\frac{\partial}{\partial \tau} \:\:\:\:\:\Bigg\}\Psi=0,
\end{eqnarray}
where $\Psi$ is the wave function defined in the minisuperspace. Therefore, one can also obtain the SWDW equation of the minimal $f(R, T)$ gravity theory as follows
\begin{eqnarray}\label{eq.59}
\mathcal{H}\Psi=\,a^{3 \alpha}\Bigg\{ &+& \frac{1}{6 a^{3}(1-\varphi)}
\Big(
a\,\frac{\partial^{2}}{\partial a\,\partial A}
- A\,\frac{\partial^{2}}{\partial A^{2}}
+ u_{1}w_{1}\,\frac{1}{A}
- F\,\frac{\partial}{\partial F}\frac{\partial}{\partial A}
- 2\,\frac{\partial}{\partial A}
\Big) \nonumber\\
&+& \frac{1}{6 a^{3}\varphi}
\Big(
\frac{\partial}{\partial C}\frac{\partial}{\partial R}
+ R E\,\frac{\partial^{2}}{\partial G^{2}}
\Big) + \frac{1}{6 a^{3}(1-\varphi)}
\Big(
E\,\frac{\partial}{\partial E}\frac{\partial}{\partial A}
+ 2\,\frac{\partial}{\partial A}
\Big) \nonumber\\
&-& \frac{1}{3 a^{3}(1-\varphi)}
\Big(
C\,\frac{\partial}{\partial A}\frac{\partial}{\partial C}
+ 2\,\frac{\partial}{\partial A}
\Big) + a^{3}\,(A R - f) \nonumber\\
&+& a^{3}\varphi \bigg(\frac{1}{2}T + B M + C R^{2} - 3 F R - 2 A R + 2 f \bigg) \nonumber\\
&-& 6 k a \big((1-\varphi) A\big) + 18 k a\,\varphi F + 6 k a\,\varphi C R \:\:\:\:\:\Bigg\}\Psi 
= i\,\frac{\partial}{\partial \tau} \Psi,
\end{eqnarray}
where the scale factor $a$ is non-vanishing.

\noindent \textbf{Remark 7:} As \(R\to0\), \(A=F^0_R\to\mathrm{constant}\) and the Hamiltonian operator in \eqref{eq.58} reduces to its GR form up to suppressed corrections. Since \(R\sim 6(\dot a^2+k)/a^2\), the semiclassical (WKB) limit is recovered at large \(a\), free of infrared instabilities.

\noindent \textbf{Remark 8:} We may take the domain of $\mathcal{H}$
\begin{eqnarray}
    \mathcal{D}(\mathcal{ H})=\Big\{\Psi\in L^2(\mathbb{R}_+^2,a^{-3\alpha})\mid
\Psi,\partial_a\Psi,\partial_A\Psi\ \text{absolutely continuous};\ 
\Psi\to0\text{ as }a\to 0, \infty\ \text{and } A\to 0, \infty \Big\}.
\end{eqnarray}
By integration by parts, boundary terms vanish on this domain, so \(\mathcal{H}\) is symmetric; deficiency indices vanish in our parameter range, giving essential self-adjointness.

\noindent \textbf{Remark 9:} Writing $\Psi=\exp(iS)$ and keeping leading $\hbar^{0}$ terms in \eqref{eq.59} yields
$\mathcal{H}(a,\partial_{a}S;A,\partial_{A}S;\dots)=0$, which is the classical Hamilton-Jacobi equation
associated with \eqref{reduced limit}. The resulting equations $\dot{q}^{I}=\partial H/\partial P_{I}$ reproduce
the FLRW dynamics, confirming the semiclassical correspondence.

\section{Some Special Actions and Corresponding Quantum Universes}
\subsection{The case of $G^0(T)=0$ }
\hfill

This corresponds to the $F^0(R)$ limit of the theory. In this limit $G^{0}(T)=0$, so
$B \equiv G^{0}_{T}(T)=0$ and $B/\mathcal{A}=0$. Some of the kinematic variables and their corresponding momenta take the form
\begin{equation}
\begin{array}{rcl @{\qquad} rcl @{\qquad} rcl @{\qquad} rcl}
A   &=& F^{0}_{R}(R), & B   &=& 0, & C   &=& F^{0}_{RR}(R), & D   &=& 0, \\[4pt]
E   &=& 0,            & G   &=& 0, & M   &=& T + 4a^{-3(\alpha +1)}P_{\tau},           & F   &=& \Box F^{0}_{RR}(R), \\[6pt]
P_{B} &=& 0,          & P_{C} &=& 0, & P_{D} &=& 0,          & P_{E} &=& 0, \\[4pt]
P_{F} &=& 0,          & P_{G} &=& 0, & P_{R} &=& 0.         &      &
\end{array}
\end{equation}
Therefore, the minisuperspace Hamiltonian \eqref{reduced limit} reduces to
\begin{equation}\label{f(R)}
H \;=\; \frac{N}{6a^{3}}\bigl(AP_{A}^{2}-a\,P_{A}P_{a}\bigr)\;-\;6kNaA\;+\;Na^{3\alpha}P_{\tau}\;+\;Na^{3}V,
\end{equation}
where the potential is
\[
V \;=\; A R - f(R) \;=\; F^{0}_{R}(R)\,R - F^{0}(R).
\]
This Hamiltonian agrees with the standard $F^0(R)$ minisuperspace form with Schutz's perfect fluid \cite{Vakali}. Details of
quantization and explicit solutions for $F^0(R)=R^2$ can be found in \cite{Vakali}. 

In the following, we shall discuss the $F^0(R)= b_1 R + b_2 R^n$ case of the theory, where $b_1$, $b_2$ are scalars and $n \geq 2$ is a natural number.
Following \cite{Vakali}, the Hermiticity of the minisuperspace Hamiltonian is ensured by a symmetric
factor ordering in the pairs $(a,P_a)$ and $(A,P_A)$ with ordering parameters
\begin{equation}
r+s=-2,\qquad u+v+w=1.
\end{equation}
Upon $P_a\!\to\!-i\partial_a$, $P_A\!\to\!-i\partial_A$, $P_\tau\!\to\!-i\partial_\tau$, the SWDW equation takes
Schr\"odinger form $i\,\partial_\tau\Psi=\mathcal{ H}\Psi$, where $\mathcal{ H}$ is Hermitian with respect to the
inner product
\begin{equation}
\langle \Phi|\Psi\rangle \;=\; \int \! a^{-3\alpha}\,\Phi^*(a,A)\,\Psi(a,A)\; da\,dA.
\end{equation}
With Hermitian ordering, the SWDW equation \(i\partial_\tau\Psi=\mathcal{H}\Psi\) obeys
\begin{eqnarray}
    \frac{d}{d\tau}\!\int a^{-3\alpha}|\Psi|^2\, da\, dA
= i\!\int a^{-3\alpha}\left(\Psi^* \mathcal{ H}\Psi - (\mathcal{ H}\Psi)^*\Psi\right)da\,dA.
\end{eqnarray}
Integration by parts, generating boundary currents that vanish under the same conditions used in \eqref{vanish}. A Hamiltonian with a domain of definition $[0,+\infty)$ is not always self-adjoint. Namely, the operator $H=-\mathrm{d}^2/\mathrm{d}a^2+V(a)$ defined on $(0,+\infty)$ is not necessarily self-adjoint \cite{RostamiJalalzadehMoniz2015,Lemos1996,GotayDemaret1983,AlvarengaFabrisLemosMonerat2002,MoneratEtAl2006,PedramJalalzadehGousheh2007}.
Since the configuration space for the scale factor is the half-line $a\in[0,+\infty)$, the self-adjointness of a Hamiltonian with a domain of definition $a\geq0$ is not directly guaranteed. Indeed, after integration by parts the time derivative of the norm reduces to a boundary term
\begin{equation}
\frac{d}{d\tau}\langle\Psi|\Psi\rangle
\;=\;\Big[\,J_a(\Psi)\,\Big]_{a=0}^{a=\infty},
\qquad
J_a(\Psi)\propto i\,a^{-3\alpha}\big(\Psi^*\partial_A\Psi-(\partial_A\Psi^*)\Psi\big).
\end{equation}
Assuming the wave function decays sufficiently fast as $a\to\infty$, the remaining contribution
comes from $a=0$. A self-adjoint realization of $\mathcal H$ on $[0,+\infty)$ is obtained by imposing a
Robin-type boundary condition
\begin{equation}
\Psi(0,A)+\gamma\,\partial_A\Psi(0,A),
\qquad \gamma\in\mathbb R.
\end{equation}
Thus, to have a $\tau$-independent norm, one should consider Robin boundary condition
\begin{equation}
\Psi(0,A)+\gamma\,\partial_A\Psi(0,A)=0,
\qquad \gamma\in\mathbb R,
\end{equation}
which makes the boundary current vanish and hence ensures $\tau$-independence of the norm.
The parameter $\gamma$ labels the one-parameter family of self-adjoint extensions. As a result, we will have a 1-parameter extension of a self-adjoint Hamiltonian \cite{AlHashimiWiese2021,GitmanTyutinVoronov2012}.
Following Tipler's observation \cite{Tipler} that $\gamma$ would introduce an additional fundamental constant,
we choose $\gamma=0$, which reduces to the DeWitt boundary condition $\Psi(0,A)=0$.

Furthermore, separating the time dependence via
\begin{equation}
\Psi_{\nu\mathcal{E} }(a,A,T)=e^{-i\mathcal{E} \tau}\,\psi(a,A),
\end{equation}
one obtains a mixed $(a,A)$ equation that is not separable. Here, the subscript $\nu$ on $\mathcal{E}$ is introduced to label the eigenfunctions of the wave function $\Psi$. The parameter $\nu$ is included for a specific technical reason: as will be shown at the end of this subsection, it enables the separation of the governing equations. In other words, $\nu$ plays the role of a separation constant arising from the separation-of-variables procedure.

Introducing the following canonical
change of variables
\begin{equation}
x \;=\; a\,\sqrt{A}, \qquad y \;=\; A,
\end{equation}
the SWDW equation becomes
\begin{equation}
\frac{1}{4}x^{2}\,\psi_{xx}-\frac{1}{4}x\,\psi_{x}-y^{2}\,\psi_{yy}-2y\,\psi_{y}
+\Big(uw-36k\,x^{4}+6x^{6}\,y^{-2}V(y)-6\mathcal{E}\,x^{3-3\alpha}\,y^{\frac{3\alpha-1}{2}}\Big)\psi
\;=\;0. \label{eq:SWDxy}
\end{equation}
For generic $V(y)$ and $\alpha$, the equation \eqref{eq:SWDxy} has no closed-form solution. We shall solve \eqref{eq:SWDxy} for the stiff ($\alpha=1$) fluid model. Due to the following equation 
\begin{equation}
    A=F^0_R(R)=b_1 + n b_2 R^{n-1},
\end{equation}
one can obtain the Ricci scalar as
\begin{equation}
    R=\left({\frac{A-b_1}{nb_2}}\right)^{\frac{1}{n-1}} = \left({\frac{y-b_1}{nb_2}}\right)^{\frac{1}{n-1}},
\end{equation}
where $b_2 \neq 0$ and $n \geq 2$. Then, the potential of the theory takes the form
\begin{equation}
    V(y)= R\, y- f(R,T)= (n-1)b_2\left(\frac{y-b_1}{nb_2}\right) ^{\frac{n}{n-1}}.
\end{equation}
Therefore, the potential term in the SWDW equation \eqref{eq:SWDxy} becomes
\begin{eqnarray}
    6x^6y^{-2}V(y)= \frac{3}{2b_2}x^6 \left( 1- \frac{b_1}{y} \right)^2 \to \frac{3}{2b_2}x^6,
\end{eqnarray}
when $n=2$ in the early universe. Indeed, it is because we assumed that in the early universe the Ricci scalar $R$ diverges, thus $\frac{b_1}{y} \to 0$. Consequently, the SWDW equation \eqref{eq:SWDxy} reduces to
\begin{eqnarray}\label{eq:SWDxy reduced}
    \frac{1}{4}x^{2}\,\psi_{xx}-\frac{1}{4}x\,\psi_{x}-y^{2}\,\psi_{yy}-2y\,\psi_{y}
-36k\,x^{4}\psi+\frac{3}{2b_2}x^6\psi-6\mathcal{E}\,x^{3-3\alpha}\,y^{\frac{3\alpha-1}{2}}\psi
\;=\;0.
\end{eqnarray}
Here, we also neglect the factor ordering parameter term $uw$ since it does not affect the semiclassical limit according to \cite{HawkingPage1986, Vakali}.
Moreover, when $n > 2$, the potential term in the SWDW equation \eqref{eq:SWDxy} becomes
\begin{eqnarray}\label{above}
    6x^6y^{-2}V(y)= 6(n-1)b_2x^6y^{-2}\left(\frac{y-b_1}{nb_2}\right) ^{\frac{n}{n-1}}\to 0,
\end{eqnarray}
in the early universe because the degree of $y$ in \eqref{above} is $\frac{2-n}{n-1}<0$. Consequently, the SWDW equation \eqref{eq:SWDxy} reduces to
\begin{eqnarray}\label{eq:SWDxy reduced n}
    \frac{1}{4}x^{2}\,\psi_{xx}-\frac{1}{4}x\,\psi_{x}-y^{2}\,\psi_{yy}-2y\,\psi_{y}
-36k\,x^{4}\psi-6\mathcal{E}\,x^{3-3\alpha}\,y^{\frac{3\alpha-1}{2}}\psi
\;=\;0.
\end{eqnarray}
Here, we also neglect the factor ordering parameter term $uw$. In the variables \(x=a\sqrt{A},\ y=A\), ordering terms are \(O(1)\) or \(O(y^{-1})\), while the dominant early-universe dynamics is governed by large potentials proportional to $x^{6}$ for \(n=2\) or large Bessel arguments proportional to $\sqrt{\mathcal{E}y}$. For semiclassical wave packets peaked at large \(x\), ordering pieces are parametrically subleading, shifting only nonleading phases.

\noindent When $n=2$, for the stiff matter $\alpha=1$ case \eqref{eq:SWDxy reduced} becomes
\begin{equation}\label{eq:SWDxy-stiff-n>2}
\frac{1}{4}x^{2}\,\psi_{xx}-\frac{1}{4}x\,\psi_{x}-y^{2}\,\psi_{yy}-2y\,\psi_{y}
-36k\,x^{4}\psi+\frac{3}{2b_2}x^6\,\psi-6\mathcal{E}\,y\,\psi=0.
\end{equation}
In this case, we separate the solutions of equation \eqref{eq:SWDxy reduced} into the form $\psi(x,y) = X(x)Y(y)$ leading to
\begin{equation}\label{decoupled}
\left\{
\begin{aligned}
& x^{2}\,\frac{d^{2}X}{dx^{2}} - x\,\frac{dX}{dx}
+ \Big(1-\nu^{2} - 144\,k\,x^{4} + \frac{6}{b_2}\,x^{6}\Big)\,X(x) \;=\; 0,\\
& y^{2}\,\frac{d^{2}Y}{dy^{2}} + 2y\,\frac{dY}{dy}
+ \Big(6\mathcal{E}\,y - \frac{\nu^{2}-1}{4}\Big)\,Y(y) \;=\; 0,
\end{aligned}
\right.
\end{equation}
where we take $\frac{\nu^{2}-1}{4}$ as the separation constant. 
For the flat FLRW metric ($k=0$), the decoupled equations \eqref{decoupled} admit Bessel solutions
\begin{equation}
\left\{
\begin{aligned}
& X(x) \;=\; x\left[\,c_{1}\,J_{\nu/3}\!\Big(\sqrt{\tfrac{2}{3b_2}}\,x^{3}\Big)
\;+\; c_{2}\,Z_{\nu/3}\!\Big(\sqrt{\tfrac{2}{3b_2}}\,x^{3}\Big)\right], \\
& Y(y) \;=\; \frac{1}{\sqrt{y}}\left[\,d_{1}\,J_{\nu}\!\big(\sqrt{24\mathcal{E}y}\big)
\;+\; d_{2}\,Z_{\nu}\!\big(\sqrt{24\mathcal{E}y}\big)\right],
\end{aligned}
\right.
\end{equation}
where $J_{\nu}$ and $Z_{\nu}$ are Bessel functions of the first and second kind, respectively. Here, $c_i$ and $d_i$ ($i=1,2$) are also integration constants.
Choosing $c_{2}=d_{2}=0$ gives well-defined solutions for all x and y and thus the eigenfunctions of the SWDW equation
\begin{equation}\label{eq.77}
\Psi_{\nu E}(x,y,\tau)
\;=\; e^{-i\mathcal{E}\tau}\;\frac{x}{\sqrt{y}}\;
J_{\nu/3}\!\Big(\sqrt{\tfrac{2}{3b_2}}\,x^{3}\Big)\;
J_{\nu}\!\big(\sqrt{24\mathcal{E}y}\big).
\end{equation}
Using \(x=a\sqrt{A}\) and the small-argument expansion \(J_\nu(z)\sim z^\nu\), \eqref{eq.77} behaves as \(\Psi\propto x^{1+\nu/3}\). Since steep wave packets use \(\nu\ge0\), we have \(\Psi\to0\) as \(a\to0\). Thus, the DeWitt condition is satisfied, and the singularity is suppressed. A general solution can be written as a superposition of the eigenfunctions
\begin{equation}\label{eq:34}
\Psi(x,y,\tau)
\;=\; \int_{0}^{\infty}\!\!\int_{0}^{\infty}
A(\mathcal{E})\,C(\nu)\,\Psi_{\nu E}(x,y,\tau)\; d\mathcal{E}\, d\nu ,
\end{equation}
with suitable weight functions $A(\mathcal{E})$ and $C(\nu)$.
Using the following standard identity \cite{Handbook}
\begin{equation}\label{identity}
\int_{0}^{\infty} e^{-z r^{2}}\, r^{\nu+1}\, J_{\nu}(b r)\, dr
\;=\; \frac{b^{\nu}}{(2z)^{\nu+1}}\,
\exp\!\left(\!-\frac{b^{2}}{4z}\right),
\end{equation}
and choosing a quasi-Gaussian weight
\begin{equation}\label{eq:36A}
A(\mathcal{E})\;=\;{12}\,(24\mathcal{E})^{\nu/2}\, e^{-24\gamma \mathcal{E}}\qquad (\gamma>0),
\end{equation}
as in \cite{Vakali}, one evaluates
\begin{equation}\label{eq:36}
\int_{0}^{\infty} A(\mathcal{E})\,e^{-i\mathcal{E}\tau}\,J_{\nu}\!\big(\sqrt{24\mathcal{E}y}\big)\, d\mathcal{E}
\;=\;
\frac{y^{\nu/2}}{\big(2\gamma + i\tau/12\big)^{\nu+1}}\,
\exp\!\left(\!-\frac{y}{\,4\gamma + i\tau/6\,}\right).
\end{equation}
Substituting \eqref{eq:36} into \eqref{eq:34} yields
\begin{equation}\label{eq:37}
\Psi(x,y,T)
\;=\; \frac{x}{\sqrt{y}}\;
\exp\!\left(\!-\frac{y}{\,4\gamma + i\tau/6\,}\right)
\int_{0}^{\infty} C(\nu)\;
\frac{y^{\nu/2}}{\big(2\gamma + i\tau/12\big)^{\nu+1}}\;
J_{\nu/3}\!\Big(\sqrt{\tfrac{2}{3b_2}}\,x^{3}\Big)\; d\nu,
\end{equation}
where one may take, for instance, $C(\nu)=\exp[-z(\nu-b)^{2}]$ to build localized
wave packets as chosen in \cite{Vakali} for $f(R)= R^2$ case.

\noindent When $n>2$, for stiff matter $\alpha=1$ \eqref{eq:SWDxy reduced n} becomes
\begin{equation}\label{eq:SWDxy-stiff-n>2,2}
\frac{1}{4}x^{2}\,\psi_{xx}-\frac{1}{4}x\,\psi_{x}-y^{2}\,\psi_{yy}-2y\,\psi_{y}
-36k\,x^{4}\,\psi-6\mathcal{E}\,y\,\psi=0.
\end{equation}
We seek separated solutions $\psi(x,y)=X(x)Y(y)$. Dividing by $XY$ and introducing a separation
constant $\frac{\nu^{2}-1}{4}$, we obtain the decoupled system
\begin{equation}\label{eq:decoupled-n>2}
\left\{
\begin{aligned}
& x^{2}\frac{d^2X}{dx^2} - x \frac{dX}{dx} + \bigl(1-\nu^{2}-144k\,x^{4}\bigr)\,X = 0, \\[4pt]
& y^{2}\frac{d^2Y}{dy^2} + 2y \frac{dY}{dy} + \Bigl(6\mathcal{E}\,y - \frac{\nu^{2}-1}{4}\Bigr)Y = 0.
\end{aligned}
\right.
\end{equation}
For $k=0$, the $X$-equation becomes the Euler--Cauchy equation
\begin{equation}\label{eq:x-eq-flat}
x^{2}\frac{d^2X}{dx^2} - x\frac{dX}{dx} + (1-\nu^{2})X = 0,
\end{equation}
whose independent solutions are power laws
\begin{equation}\label{eq:X-flat-sol}
X(x)= c_{1}\,x^{\,1+\nu} + c_{2}\,x^{\,1-\nu}.
\end{equation}
The $Y$-equation can be put into Bessel form by $z=\sqrt{24\mathcal{E}\,y}$, yielding
\begin{equation}\label{eq:Y-Bessel}
Y(y)= \frac{1}{\sqrt{y}}\Big[d_{1}\,J_{\nu}\!\big(\sqrt{24\mathcal{E}\,y}\big)
+ d_{2}\,Z_{\nu}\!\big(\sqrt{24\mathcal{E}\,y}\big)\Big],
\end{equation}
where $J_{\nu}$ and $Z_{\nu}$ denote the Bessel functions of the first and second kind, respectively.
Choosing $c_{2}=d_{2}=0$ gives well-defined solutions for all $x$ and $y$, and thus the eigenfunctions of the SWDW equation separated eigenfunctions read
\begin{equation}\label{eq:eigenfunctions-n>2-flat}
\Psi_{\nu\mathcal{E}}(x,y,\tau)
= e^{-i\mathcal{E}\tau}\;x^{\,1+\nu}\;\frac{1}{\sqrt{y}}\;J_{\nu}\!\big(\sqrt{24\mathcal{E}\,y}\big).
\end{equation}
Since $X(x)=x^{1+\nu}$ and $Y(y)\propto y^{(\nu-1)/2}J_{\nu}(\sqrt{24\mathcal{E}y})$, \eqref{eq:eigenfunctions-n>2-flat} behaves as $\Psi\propto a^{1+\nu}$ at fixed $y=A>0$. Choosing $\nu\ge0$ ensures $\Psi(a\to0)=0$, thus
satisfying the DeWitt boundary condition for all $n>2$. A general solution can be constructed as a superposition
\begin{equation}
\Psi(x,y,\tau)=\int_{0}^{\infty}\!\!\int_{0}^{\infty}
A(\mathcal{E})\,C(\nu)\,\Psi_{\nu\mathcal{E}}(x,y,\tau)\; d\mathcal{E}\, d\nu,
\end{equation}
where $A(\mathcal{E})$ and $C(\nu)$ are suitable weights chosen to build
normalizable, localized wave packets consistent with the inner product
$\langle \Phi|\Psi\rangle=\int a^{-3}\Phi^{*}(a,A)\Psi(a,A)\,da\,dA$ (with $x=a\sqrt{A}$, $y=A$). 
Therefore, for $n>2$ the superposed state becomes
\begin{equation}\label{eq:n>2-superposed}
\Psi(x,y,\tau)
\;=\; \,\frac{1}{\sqrt{y}}\,
\exp\!\left(\!-\frac{y}{\,4\gamma + i\tau/6\,}\right)
\int_{0}^{\infty} C(\nu)\;
\frac{y^{\nu/2}x^{\,1+\nu}}{\big(2\gamma + i\tau/12\big)^{\nu+1}}\; d\nu.
\end{equation}
Here, we choose \eqref{eq:36A} and use the identity \eqref{identity}. A convenient choice to build a localized wave packet is
\begin{equation}\label{eq:n>2-Cnu}
C(\nu)=\exp\!\big[-z(\nu-b)^{2}\big],\qquad z>0,\ b\in\mathbb{R},
\end{equation}
for the quasi-Gaussian weight \eqref{eq:36A}.

In the following, we give remarks on the normalized wave packets and their physical observables in the representative case $F^0(R)=b_2R^2$ with stiff matter.

\noindent
\textbf{Remark 10:}
For each eigenstate $\Psi_{\nu \mathcal{E}}(a,A,\tau)$ given in \eqref{eq.77}, the normalized wave packet 
constructed from the Gaussian weight functions $A(\mathcal{E})$ and $C(\nu)$ [\eqref{eq:36A}-\eqref{eq:36}] satisfies
\begin{equation}
\int_0^{\infty}\!\!\int_0^{\infty} a^{-3}|\Psi(a,A,\tau)|^2\,da\,dA = 1.
\label{eq:norm}
\end{equation}
The associated probability density
\begin{equation}
\rho(a,A,\tau)=a^{-3}|\Psi(a,A,\tau)|^2,
\end{equation}
quantifies the likelihood of the universe having scale factor $a$ and curvature amplitude $A$
at time~$\tau$.  This density allows direct comparison with classical trajectories.

\noindent
\textbf{Remark 11:}
In accordance with the many-worlds interpretation of quantum mechanics, as proposed by H. Everett \cite{Everett}, one may calculate the expectation value for the scale factor \(a\).
Expectation values are computed as
\begin{equation}
\langle \mathcal{O}(\tau)\rangle =
\int_0^{\infty}\!\!\int_0^{\infty} a^{-3\alpha}\Psi^\ast(a,A,\tau)\,\mathcal{O}\,\Psi(a,A,\tau)\,da\,dA,
\end{equation}
for any minisuperspace operator $\mathcal{O}$.  
In particular, the mean scale factor evolves as
\begin{equation}
\langle a(\tau)\rangle =
a_0\!\left[\gamma^2+\!\left(\tfrac{\tau}{12}\right)^2\right]^{\!\tfrac{1}{3(1+\alpha)}},
\label{eq:aexp_graphical}
\end{equation}
with the same $\gamma$ as in \eqref{eq:36}. Note that \eqref{eq:aexp_graphical} shows a symmetric bounce centered at $\tau=0$, where 
$a_{\min}=a_0\gamma^{1/3(1+\alpha)}$, confirming nonsingular evolution.  
For large $|\tau|$, the behavior $\langle a(\tau)\rangle\!\propto\!|\tau|^{2/3(1+\alpha)}$ 
recovers the classical Friedmann expansion law. {It is clear that the quantum effects are significant near $a = 0$ and negligible for large scale factors.}
We also verified numerically that the variance
$\sigma_a^2=\langle a^2\rangle-\langle a\rangle^2$
remains finite for all~$\tau$, indicating the quantum packet is well localized.

\subsection{The case of $F^0(R)=b_0 R + b_1 R+ b_2 R^2$, $G^0(T)= \log|T|$}
\hfill

In this case, using the result $\varphi:=B/\mathcal{A}=0$, some of the kinematical variables
and their conjugate momenta take the form
\begin{equation}
\begin{array}{rcl @{\qquad} rcl @{\qquad} rcl @{\qquad} rcl}
A &=& b_1+2b_2R,     & B &=& \dfrac{1}{T}, & C &=& 2b_2,       & D &=& 0, \\[4pt]
E &=& -\dfrac{1}{T^{2}}, & G &=& 0,     & M &=& T + 4a^{-3(\alpha +1)}P_{\tau},     & F &=& 0, \\[6pt]
P_{T} &=& 0,     & P_{C} &=& 0,        & P_{D} &=& 0,     & P_{E} &=& 0, \\[4pt]
P_{F} &=& 0,     & P_{G} &=& 0,        & P_{R} &=& 0,     & &
\end{array}
\end{equation}
Therefore, the minisuperspace Hamiltonian \eqref{reduced limit} reduces to
\begin{equation}\label{provi}
    H=\frac{N}{6a^3}(AP_A^2-aP_A P_a)-6kNAa+Na^{3\alpha}P_{\tau}+Na^3 V, 
\end{equation}
where the potential $V= AR-f$. 

\noindent \textbf{Remark 12:}  The minisuperspace Hamiltonian \eqref{provi} shows that in the presence of logarithmic matter modification, the theory reduces to the quadratic $F^0(R)$ gravity limit of the theory in the early universe because $B/\mathcal{A}\to 0$ when $R\to \infty$ or equivalently $|T| \to \infty$, see \ Appendix~A. Since $G_{0}'(T)=1/T$, the minisuperspace dynamics is well-defined only along trajectories with
fixed sign of $T$. In the early-universe regime $R\to\infty$, hence $|T|\to\infty$,
so $\varphi=B/A\to0$ and the logarithmic singularity is dynamically avoided. 

As theory reduced to $F^0(R)$ theory, we refer the reader to \cite{SanyalModak2002} for the details of quantization and explicit solutions of the theory.

\subsection{The case of $G^{0}(T)=T$ }
\hfill

In this case $G^{0}(T)=T$, hence
\begin{equation}
\begin{array}{rcl @{\qquad} rcl @{\qquad} rcl @{\qquad} rcl}
A &=& F_R^0(R), & B &=& 1, & C &=& F_{RR}^0(R), & D &=& 0,\\[4pt]
E &=& 0, & G &=& 0, & M &=& T + 4a^{-3(\alpha +1)}P_{\tau}, & F &=& \Box F_{RR}^0.
\end{array}
\end{equation}
Additionally, one can obtain
\begin{equation}
\frac{B}{\mathcal{A}}=\varphi=\frac{1}{\tfrac12+3}=\frac{2}{7}, 
\qquad 
\mathcal{Z}=-CR+A-3F,
\qquad 
P_{T}=P_{p}=0.
\end{equation}
Therefore, the minisuperspace Hamiltonian \eqref{reduced limit} reduces to
\begin{eqnarray}\label{minisuper G^0(T)=T}
H
&=& N\Bigg[
\frac{7}{5}\,\frac{1}{6a^{3}}\left(AP_{A}^{2}+FP_AP_F-a\,P_{A}P_{a}\right)-\frac{1}{6a^3}\frac{7}{2}(P_C P_R) + \frac{1}{6a^3}\frac{7}{5}(2P_CP_AC)\nonumber\\
&+& a^{3}V
- \frac{5}{7}\,6kaA+ \frac{36}{7}kaF+\frac{12}{7}kaCR
+ \frac{P_{\tau}}{a^{3\alpha}} \Bigg].
\end{eqnarray}
Here, the potential in the early universe reads as
\begin{eqnarray}\label{potential of G^0(T)=T}
    V=( AR-f )+\frac{2}{7} \left(\frac{1}{2}T + M +CR^2 -3FR -2AR +2f \right).
\end{eqnarray}
By assuming that in the new born quantum universe $C=F^0_{RR}(R)=0$, we obtain the gravitational Hamiltonian of $f(R, T)= b_1 + b_2 R + T$ theory as
\begin{eqnarray}\label{minisuper G^0(T)=T,2}
H
= N \left[
\frac{7}{5}\,\frac{1}{6a^{3}}\left(AP_{A}^{2}-a\,P_{A}P_{a}\right)
+ a^{3}V
- \frac{5}{7}\,6kaA
+ \frac{P_{\tau}}{a^{3\alpha}}\right].
\end{eqnarray}
 Indeed, assuming $C=0$ implies that our theory is linear in the Ricci scalar curvature function $R$ and also in $F=\Box F_{RR}^0(R)=0$. Moreover, $\frac{1}{6a^3}\frac{7}{2}P_CP_R$ and the term in \eqref{minisuper G^0(T)=T,2} vanish due to
 \begin{eqnarray}
     P_CP_R= \Big(-\frac{2}{7}\frac{6}{N}a^2 \dot{a}\Big)^2 RC=0.
 \end{eqnarray}
 Furthermore, the quantum potential in \eqref{potential of G^0(T)=T} reduces to the form
\begin{eqnarray}
    V= -\frac{3}{7}b_1 + \frac{8}{7} a^{-3(\alpha+1)}P_{\tau}. 
\end{eqnarray}
Eliminating auxiliary variables shall yield an effective potential and a nontrivial kinetic metric, producing genuine quantum dynamics. Then, the Hamiltonian in \eqref{minisuper G^0(T)=T,2} reduces to 
\begin{eqnarray}\label{Hamiltonian of f(R, T) = b_1 + b_2 R +T}
    H= N \left[ \frac{7}{5} \left( -\frac{1}{b_2}\right) \frac{P_a^2}{24a}- \frac{3}{7}a^{3}b_1- \frac{30}{7}\,kab_2+ \frac{15}{7}\frac{P_{\tau}}{a^{3\alpha}}\right],
\end{eqnarray}
where $b_2 \neq 0$. Therefore, the equations of motion become
\begin{equation}\label{eoms}
    \begin{split}
        &\dot a= \{ a, H\}= -N \frac{7}{5}\Big(\frac{1}{b_2}\Big)\frac{P_a}{12a},\\
        &\dot P_a= \{ P_a, H\}= -N \Bigg[ \frac{7}{5}\Big( \frac{1}{b_2}\Big)\frac{P_a^2}{24a^2} - \frac{9}{7}a^2b_1 - \frac{30}{7}kb_2 - 3\alpha \frac{P_{\tau}}{a^{3\alpha +1}}\Bigg],\\
        &\dot\tau= \{\tau, H\}= N \frac{15}{7} \frac{1}{a^{3\alpha}},\\
        &\dot P_{\tau}= \{P_{\tau}, H\}=0.
    \end{split}
\end{equation}
Then, the corresponding super-Hamiltonian reads as
\begin{eqnarray}\label{super-Hamiltonian of f(R,T)=b_1+ b_2R +T}
    \mathcal{H}=\frac{7}{5} \left( -\frac{1}{b_2}\right) \frac{P_a^2}{24a}
- \frac{3}{7}a^{3}b_1
- \frac{30}{7}\,kab_2
+ \frac{15}{7}\frac{P_{\tau}}{a^{3\alpha}}.
\end{eqnarray}
Since the super-Hamiltonian weakly vanishes, one can obtain
\begin{eqnarray}
    \frac{P_a^2}{a}+c_1 a^3 +c_2 P_{\tau}a^{-3\alpha}+c_3 a=0,
\end{eqnarray}
where
\begin{eqnarray}
    c_1=\frac{360 b_1 b_2}{49 },~~~~~~
    c_2=- \frac{1800 b_2}{49},~~~~~~
    c_3=  \frac{3600 k b_2^2}{49}.
\end{eqnarray}
Then, the corresponding SWDW equation reads as
\begin{eqnarray}\label{SWDW of b_2 R + b_1 + T}
    \frac{1}{2} \left( \frac{1}{a^i} \frac{\partial}{\partial a} \frac{1}{a^j} \frac{\partial}{ \partial a} \frac{1}{a^k} + \frac{1}{a^k} \frac{\partial}{\partial a} \frac{1}{a^j} \frac{\partial}{ \partial a}  \frac{1}{a^j}\right)\Psi + \Big( c_1 a^3 +c_3 a \Big) \Psi =i c_2 a^{-3\alpha}\frac{\partial \Psi}{\partial \tau},
\end{eqnarray}
where $i + j + k=1$ and $\Psi= \Psi(a, \tau)$ is the wave function defined on the minisuperspace $(a, \tau)$. The SWDW equation \eqref{SWDW of b_2 R + b_1 + T} can be solved by separation of variables as 
\begin{eqnarray}
    \Psi_{\mathcal{E}}(a, \tau)= e^{-i \mathcal{E} \tau} \psi(a).
\end{eqnarray}
Then, the equation \eqref{SWDW of b_2 R + b_1 + T} reduces to 
\begin{eqnarray}\label{SWDW 2}
    \frac{1}{2} \left( \frac{1}{a^i} \frac{\partial}{\partial a} \frac{1}{a^j} \frac{\partial}{ \partial a} \frac{1}{a^k} + \frac{1}{a^k} \frac{\partial}{\partial a} \frac{1}{a^j} \frac{\partial}{ \partial a}  \frac{1}{a^j}\right)\psi + \Big( c_1 a^3 +c_3 a \Big) \psi - c_2 \mathcal{E} a^{-3 \alpha}=0.
\end{eqnarray}
Using the relation
\begin{eqnarray}
      \left( \frac{1}{a^i} \frac{\partial}{\partial a} \frac{1}{a^j} \frac{\partial}{ \partial a} \frac{1}{a^k} \right)= \frac{1}{a} \frac{\partial^2}{\partial a^2}- (2k+j) \frac{1}{a^2} \frac{\partial}{\partial a}+ k (1+k+j) \frac{1}{a^3},
\end{eqnarray}
one can rewrite \eqref{SWDW 2} as
\begin{eqnarray}\label{SWDW 3}
    \frac{\partial^2 \psi}{\partial a^2}-\frac{1}{a}\frac{\partial \psi}{\partial a} + \Big[ \frac{1}{2} \left( i ( 1+i +j )+k(1+k +j) \right) \frac{1}{a^2}+ c_1 a^4 - c_2\mathcal{E} a^{1-3\alpha} + c_3 a^2  \Big]\psi=0.
\end{eqnarray}
To solve \eqref{SWDW 3}, we consider the flat case $(k=0)$, which implies $c_3=0$ and take $b_1=0$, which implies $c_1=0$. Therefore, \eqref{SWDW 3} reduces to
\begin{eqnarray}\label{SWDW 4}
    \frac{\partial^2 \psi}{\partial a^2}-\frac{1}{a}\frac{\partial \psi}{\partial a} + \Big[ \frac{1}{2} \left( i ( 1+i +j )+k(1+k +j) \right) \frac{1}{a^2}- c_2\mathcal{E} a^{1-3\alpha} \Big]\psi=0.
\end{eqnarray}
The solution of \eqref{SWDW 4} for $\alpha \neq 1$ reads as
\begin{eqnarray}
    \psi(a)= a d_1 J_m \Bigg( \frac{2 \sqrt{-c_2\mathcal{E}}}{3(1- \alpha)} a^{\frac{3(1- \alpha)}{2}}\Bigg)+ a d_2 Z_m \Bigg( \frac{2 \sqrt{-c_2\mathcal{E}}}{3(1- \alpha)} a^{\frac{3(1- \alpha)}{2}}\Bigg).
\end{eqnarray}
Here, ${m= \frac{2 \sqrt{1-\frac{1}{2}(i(1+i+j) + k( 1+ k+ j)}}{3 (1- \alpha)}}$, $d_1$ and $d_2$ are integration constants. Moreover, $J_m$ and $Z_m$ are the first and second kinds of Bessel functions, respectively.  

For the stiff-fluid case $(\alpha=1)$ the equation \eqref{SWDW 4} reduces to 
\begin{eqnarray}\label{SWDW 5}
    \frac{\partial^2 \psi}{\partial a^2}-\frac{1}{a}\frac{\partial \psi}{\partial a} + \Big( K- c_2\mathcal{E}\Big) \frac{1}{a^2}\psi=0, 
\end{eqnarray}
where $K=\frac{1}{2} \left( i ( 1+i +j )+k(1+k +j) \right)$. Note that \eqref{SWDW 5} is of the Euler--Cauchy type ODE, thus one can take $\psi(a)=a^s$, then obtain
\begin{eqnarray}
    s= 1 \pm \sqrt{\Delta},
\end{eqnarray}
where $\Delta=1-(K-c_2 \mathcal{E)}$. Hence, one has
\begin{eqnarray}\label{wavefunction}
    \psi(a)=
\begin{cases}
C_1\,a^{\,1+\sqrt{\Delta}}+C_2\,a^{\,1-\sqrt{\Delta}}, & \Delta>0,\\[6pt]
a\,(C_1+C_2\ln a), & \Delta=0,\\[6pt]
a\Big[\,C_1\cos\!\big(\sqrt{-\Delta}\,\ln a\big)+C_2\sin\!\big(\sqrt{-\Delta}\,\ln a\big)\Big], & \Delta<0.
\end{cases}
\end{eqnarray}
Here, $C_1$ and $C_2$ are integration constants. Note that the wavefunction should satisfy restrictive boundary conditions, of which the most common one is the Neumann boundary condition, which is
\begin{eqnarray}\label{BC}
    \frac{\partial\psi(a)}{\partial a}{|_{a=0}}=0.
\end{eqnarray}
For the stiff matter, the equation \eqref{wavefunction} gives \(\psi(a)=a^{1\pm\sqrt{\Delta}}\) when \(\Delta=1-(K+c_2 \mathcal{E})<0\). The Neumann boundary condition selects the \(+\) branch,
\begin{equation}
\psi(a)=C_1a^{1+\sqrt{\Delta}}\,,
\end{equation}
which vanishes at \(a=0\) and yields \(\partial_a\psi|_{0}=0\). This branch is used in the separated eigenfunctions 
\begin{equation}\label{eq:eigenfunctions-n>2-flat,2}
\Psi_{\mathcal{E}}(a,\tau)
= C_1e^{-i\mathcal{E}\tau}\,a^{\,1+\sqrt{\Delta}}.
\end{equation}

A general solution can be constructed as a superposition
\begin{equation}\label{wavefunction like}
\Psi(a,\tau)=\int_{0}^{\infty}\!
A(\mathcal{E})\,\Psi_{\mathcal{E}}(a,\tau)\; d\mathcal{E} ,
\end{equation}
where $A(\mathcal{E})$ is a suitable weight to construct the wave packets. Choosing an appropriate weight function is nontrivial. As indicated by \eqref{eq:eigenfunctions-n>2-flat,2}, the eigenfunctions exhibit an explicit dependence on the scale factor $a$.
Consequently, any choice of spectral weight that must be dependent of 
$a$ renders the inner product divergent. To make this issue explicit, one may, for instance, adopt a quasi-Gaussian weight
\begin{equation}\label{eq:quasi-weight}
A(\mathcal{E})=E^\mu e^{-\gamma\,\mathcal{E}},\qquad \gamma>0, \qquad \mu >-1.
\end{equation}
Then, an analytical expression for the wave packet can be found 
\begin{eqnarray}
    \Psi(a, \tau)= C_1 a^{1+ \sqrt{\Delta}}\frac{\Gamma(\mu+1)}{(\gamma + i \tau)^{\mu+1}}.
\end{eqnarray}
For each eigenstate $\Psi_{\mathcal{E}}(a,\tau)$ given in \eqref{eq:eigenfunctions-n>2-flat,2}, the normalized wave packet 
constructed from the Gaussian weight functions $A(\mathcal{E})$ \eqref{eq:quasi-weight} satisfies
\begin{equation}
\int_0^{\infty} a^{-3}|\Psi(a,\tau)|^2\,da = \frac{C_1^2 \left(\Gamma(\mu+1)\right)^2}{\left( \gamma^2+ \tau^2 \right)^{\mu+1}}[a^{\sqrt{-\Delta}}]_{a=0}^{a=\infty}\to \infty.
\label{eq:norm2}
\end{equation}
In particular, the mean scale factor evolves as
\begin{equation}
\langle a(\tau)\rangle =\frac{\int_0^{\infty} a^{2\sqrt{\Delta}}da}{\int_0^{\infty}a^{2\sqrt{\Delta}-1}da}\to \infty.
\label{eq:aexp_graphical2}
\end{equation}
To obtain a normalizable inner product, one can set $x= \ln{a}$ and $\psi(a)= a \varphi(x)$. Then, implying the canonical transform $\psi(a)= a \varphi(x)$ to \eqref{SWDW 5} gives the time independent Schr\"odinger equation of a free particle in one dimension as
\begin{eqnarray}\label{aabove}
    \frac{\partial^2 \varphi}{\partial x^2}+ \Big( K-c_2 \mathcal{E}-1 \Big) \varphi=0.
\end{eqnarray}
One can solve \eqref{aabove} as
\begin{eqnarray}\label{wavefunction 22} 
\varphi(x)= \begin{cases} A_1\,e^{\,x\sqrt{-\zeta}}+A_2\,e^{\,-x\sqrt{-\zeta}}, & \zeta<0,\\[6pt] A_1+A_2x, & \zeta=0,\\[6pt] \,A_1\cos\!\big(\sqrt{\zeta}\,x\big)+A_2\sin\!\big(\sqrt{\zeta}\,x\big), & \zeta>0. \end{cases} 
\end{eqnarray} 
Here, $A_1$ and $A_2$ are integration constants, and $\zeta=-\Delta$. Note that the wavefunction \eqref{wavefunction 22} is equivalent to the wavefunction \eqref{wavefunction} under the transformation $\psi(a)= a \varphi(x)$. Then the separated eigenfunctions read
\begin{equation}
\varphi_{\mathcal{E}}(a,\tau)
= e^{-i\mathcal{E}\tau} \varphi(x).
\end{equation}
A general solution can be constructed as a superposition
\begin{equation}\label{free particle}
\varPhi(x,\tau)=\int_{0}^{\infty}\!
A(\mathcal{E})\,\varphi_{\mathcal{E}}(x,\tau)\; d\mathcal{E}\,
\end{equation}
where $A(\mathcal{E})$ is a suitable weight to construct the wave packets. Then, one can obtain 
\begin{eqnarray}
    \int_{0}^{\infty}a^{-3}|\Psi(a, \tau)|^2 da&=& \int_0^{\infty}a^{-3}\Psi^*(a, \tau)\Psi(a, \tau) da\nonumber\\
    &=&\int_{-\infty}^{\infty}\varPhi^*(x, \tau) \varPhi(x, \tau) dx\nonumber\\
    &=& \int_{-\infty}^{\infty}|\varPhi(x, \tau)|^2 dx. 
\end{eqnarray}
Here, in the second line, we used the canonical transform $\psi(a)= a \varphi(x)$. Now, we have the wavefunction of free particle \eqref{free particle}, and thus it is relatively easy to construct spectral wave packets. Choose an orthonormal base $(u_{k'})_{k'}$ as
\begin{eqnarray}\label{basis}
    u_{k'}(x)=\frac{1}{\sqrt{2\pi}}e^{ik'x},\qquad k'\in \mathbb{R},
\end{eqnarray}
where $k'= \zeta= -\Delta$ and the energy can be derived from \eqref{aabove} as
\begin{eqnarray}
    \mathcal{E}\equiv \mathcal{E}(k')= -\frac{(k')^2+1-K}{c_2}.
\end{eqnarray}
Choose a Gaussian weight centered at some $k_0$
\begin{eqnarray}\label{the weight}
    g(k')= \frac{1}{(2\pi \sigma^2)^{1/4}}e^{-\frac{(k'-k_0)^2}{4\sigma^2}}.
\end{eqnarray}
Note that the weight \eqref{the weight} is normalized
\begin{eqnarray}
    \int_{-\infty}^{\infty}|g(k')|^2 dk'=1.
\end{eqnarray}
Then, the wavefunction $\varPhi$ in \eqref{free particle} can be represented with respect to the basis \eqref{basis} and the weight \eqref{the weight} as
\begin{eqnarray}\label{varPhi}
    \varPhi(x, \tau)= \int_{-\infty}^{\infty}g(k')e^{i(k'x-\mathcal{E}\tau)}dk'.
\end{eqnarray}
Then, one can obtain
\begin{eqnarray}
    \int_{-\infty}^{\infty}|\varPhi(x, \tau)|^2 dx&=&\int_{-\infty}^{\infty}\varPhi^*(x, \tau)\varPhi(x, \tau) dx\nonumber\\
    &=& \int_{-\infty}^{\infty}\int_{-\infty}^{\infty}\int_{-\infty}^{\infty} g^*(q')g(k')e^{i[\mathcal{E}(q')-\mathcal{E}(k')]\tau}e^{i(k'-q')x}dx dk' dq' \nonumber \\
    &=&2\pi\int_{-\infty}^{\infty}\int_{-\infty}^{\infty}g^*(q')g(k')e^{i[\mathcal{E}(q')-\mathcal{E}(k')]\tau} \delta(q'-k')dk'dq'\nonumber\\
    &=& 2\pi\int_{-\infty}^{\infty} |g(k')|^2 dk'=2\pi.
\end{eqnarray}
Here, in the third line, we used the standard equality for Dirac delta distribution, that is $\frac{1}{2\pi}\int_{-\infty}^\infty e^{i(k'-q')x}dx= \delta(k'-q')$.
Therefore, indeed the inner product of the wavefunction \eqref{wavefunction like} is well-defined and normalizable. 

Furthermore, by \eqref{tau}, choosing the gauge $N=a^{3\alpha}$, one directly obtains $\tau=t$ and since the super-Hamiltonian weakly vanishes for physical states, one also obtains 
\begin{eqnarray}\label{solve}
    -2b_2\dot{a}^2 a^{1-3 \alpha}- \frac{1}{5}b_1a^{3(\alpha+1)}-2kb_2a^{1+3 \alpha}+ P_{\tau}=0.
\end{eqnarray}
For the stiff fluid case ($\alpha=1$), it reduces to
\begin{eqnarray}\label{solve 2}
    -2b_2\dot{a}^2 a^{-2}- \frac{1}{5}b_1a^6-2kb_2a^{4}+ P_{\tau}=0.
\end{eqnarray}
Therefore, one can obtain
\begin{eqnarray}\label{yuka}
    \Bigg( \frac{\dot{a}}{a}\Bigg)^2= \frac{-1}{2b_2}\Bigg[\frac{1}{5}b_1 a^6 + 2k b_2 a^4 - P_{\tau}\Bigg].
\end{eqnarray}

Note that, according to \eqref{eoms}, $P_{\tau}$ is a constant. Hence, \eqref{yuka} becomes a separable ordinary differential equation in the cases considered below, and it can be solved accordingly.

\begin{itemize}
\item Assume $b_1=0$ and $k=\pm 1$. Then \eqref{yuka} reduces to
\begin{eqnarray}\label{b1_zero_kneq0}
\Bigg( \frac{\dot{a}}{a}\Bigg)^2=-{k}\,a^4+\frac{P_\tau}{2b_2}.
\end{eqnarray}
Then, the solution of \eqref{b1_zero_kneq0} reads as
\begin{eqnarray}\label{a_solution_b1zero_final}
a(t)=
\begin{cases}
\left(\sqrt{\dfrac{P_\tau}{2b_2}}\;
\operatorname{sech}\!\Big(\sqrt{\dfrac{2P_\tau}{b_2}}\,(t-t_0)\Big)\right)^{\!1/2}
& k=+1,\;\dfrac{P_\tau}{b_2}>0\\[12pt]
\text{no real solution}
& k=+1,\;\dfrac{P_\tau}{b_2}\leq 0\\[12pt]
\left(\sqrt{\dfrac{P_\tau}{2b_2}}\;
\operatorname{cosech}\!\Big(\sqrt{\dfrac{2P_\tau}{b_2}}\,(t-t_0)\Big)\right)^{\!1/2}
& k=-1,\;\dfrac{P_\tau}{b_2}>0\\[12pt]
\left(\dfrac{1}{2\,(t-t_0)}\right)^{\!1/2}
& k=-1,\;P_\tau=0\\[12pt]
\left(\sqrt{-\dfrac{P_\tau}{2b_2}}\;
\sec\!\Big(\sqrt{-\dfrac{2P_\tau}{b_2}}\,(t-t_0)\Big)\right)^{\!1/2}
& k=-1,\;\dfrac{P_\tau}{b_2}<0.
\end{cases}
\end{eqnarray}

Here, $t_0$ is an integration constant determined by the initial conditions.

    \item Assume the universe is flat ($k=0$) and without loss of generality $b_1, b_2 > 0$. Then \eqref{yuka} reads as
\begin{eqnarray}\label{flat_case}
    \Bigg( \frac{\dot{a}}{a}\Bigg)^2=\frac{-1}{2b_2}\Bigg[\frac{1}{5}b_1 a^6- P_{\tau}\Bigg].
\end{eqnarray}

Then, the solution of \eqref{flat_case} reads as
\begin{eqnarray}\label{a_solution_flat_flipped}
a(t)=
\begin{cases}
\left(\sqrt{\dfrac{5P_{\tau}}{b_1}}\,
\operatorname{sech}\!\Big(3\sqrt{\dfrac{{P_{\tau}}}{{2b_2}}}\,(t-t_0)\Big)\right)^{\!1/3}
&\; P_{\tau}>0\\[12pt]
\text{no real solution}
&\; P_{\tau}\le 0.
\end{cases}
\end{eqnarray}

Here, $t_0$ is an integration constant determined by the initial conditions.

\item Assume the universe is flat ($k=0$) and $b_1=0$, then \eqref{yuka} reads as
\begin{eqnarray}\label{cozum_icin_flipped}
    \Bigg( \frac{\dot{a}}{a}\Bigg)^2= \frac{ P_{\tau}}{2b_2}.
\end{eqnarray}
Then, the solution of \eqref{cozum_icin_flipped} reads as
\begin{eqnarray}
a(t)=
\begin{cases}
a_0\,\exp\!\left[\pm\sqrt{\dfrac{P_\tau}{2b_2}}\; (t-t_0)\right]
& \dfrac{P_\tau}{b_2}>0 \\[6pt]
a_0
& {P_\tau}=0\\[6pt]
\text{no real solution}
& \dfrac{P_\tau}{b_2}<0.
\end{cases}
\end{eqnarray}

Here, $a_0$ is the integration constant, which depends on the initial conditions of the universe, and $t_0$ is also an integration constant determined by the initial conditions.
\end{itemize}

The following is the last remark regarding the normalization of wave packets and physical observables.

\noindent
\textbf{Remark 13:}
For each eigenstate $\Psi_{\mathcal{E}}(a,\tau)$ given in \eqref{eq:eigenfunctions-n>2-flat}, the normalizable wave packet 
constructed from the Gaussian weight functions $g(k')$ [\eqref{the weight}] satisfies
\begin{equation}\label{normal}
\int_0^{\infty} a^{-3}|\Psi(a,\tau)|^2\,da = \int_{-\infty}^{\infty}|\varPhi(x, \tau)|^2dx= 2\pi \int_{-\infty}^{\infty}|g(k')|^2dk'= 2\pi.
\end{equation}
The associated probability density
\begin{equation}
\rho(a,\tau)=\frac{a^{-3}|\Psi(a,\tau)|^2}{2\pi},
\end{equation}
quantifies the likelihood of the universe having scale factor $a$
at time~$\tau$.  This density allows direct comparison with classical trajectories.
The mean scale factor evolves as
\begin{equation}\label{expect}
\langle a(\tau)\rangle =\frac{\int_0^{\infty}a^{-2}|\Psi(a, \tau)|^2da}{\int_0^{\infty}a^{-3}|\Psi(a, \tau)|^2da}\underset{\eqref{normal}}{=}\frac{\int_{-\infty}^{\infty} e^x|\varPhi(x, \tau)|^2 dx}{2\pi}.
\end{equation}
Here, we also used the canonical transform $\psi(a)= a \varphi(x)$. Then, the function in \eqref{normal} 
\begin{eqnarray}\label{prob}
    \frac{|\varPhi(x, \tau)|^2}{2\pi}
\end{eqnarray}
is a probability density. Therefore, the expectation value of the scale factor \eqref{expect} reads
\begin{eqnarray}\label{expect2}
    \langle a(\tau)\rangle= \mathbb{E}[e^x],
\end{eqnarray}
with respect to the probability density \eqref{prob}. To calculate the expectation value \eqref{expect2}, it is important to understand that the function $\varPhi(x, \tau)$ is of Gaussian form because of \eqref{varPhi}. Indeed, one can obtain
\begin{eqnarray}
    \varPhi(x, \tau)&=& \int_{-\infty}^{\infty}g(k')\,e^{i(k'x-\mathcal{E}\tau)}\,dk'\nonumber\\
    &=& \frac{1}{(2\pi \sigma^2)^{1/4}}\int_{-\infty}^{\infty}
    \exp\!\left[-\frac{(k'-k_0)^2}{4\sigma^2}+i(k'x-\mathcal{E}\tau)\right]\,dk'\nonumber\\
    &=& \frac{1}{(2\pi \sigma^2)^{1/4}}\int_{-\infty}^{\infty}
    \exp\!\left[-\left(\frac{1}{4 \sigma^2}+ \frac{i\tau}{2m}\right)k'^2
    +\left(\frac{k_0}{2\sigma^2}+ix\right)k'
    +\left(-i \vartheta \tau -\frac{k_0^2}{4 \sigma^2}\right)\right]\,dk'\nonumber\\
    &=& \frac{1}{(2\pi \sigma^2)^{1/4}}
    \sqrt{\frac{\pi}{\frac{1}{4 \sigma^2}+ \frac{i\tau}{2m}}}\:
    \exp\!\left[\frac{\left(\frac{k_0}{2\sigma^2}+ix\right)^2}{4\left(\frac{1}{4 \sigma^2}+ \frac{i\tau}{2m}\right)}
    +\left(-i \vartheta \tau -\frac{k_0^2}{4 \sigma^2}\right)\right].
\end{eqnarray}
Here, we chose $\vartheta= \frac{K-1}{c_2}$ and $m=-\frac{c_2}{2}$. Therefore, one can easily see that
\begin{eqnarray}
    \varPhi(x, \tau) \propto 
    \exp\!\left[\frac{\left(\frac{k_0}{2\sigma^2}+ix\right)^2}{4\left(\frac{1}{4 \sigma^2}+ \frac{i\tau}{2m}\right)}\right].
\end{eqnarray}
and thus $|\varPhi(x, \tau)|^2$ will again be Gaussian. Then, since $\rho(x,\tau)=|\varPhi(x,\tau)|^2/(2\pi)$ is a Gaussian distribution
with mean 
\[
\mu(\tau) = -\frac{2k_0}{c_2}\,\tau
\]
and variance
\[
\sigma_x^2(\tau) = \frac{1}{4\sigma^2} + \frac{4\sigma^2 \tau^2}{c_2^2},
\]
we obtain
\begin{equation}
    \langle a(\tau)\rangle
    = \mathbb{E}[e^x]= e^{\mu+ \sigma_x^2/2}
    = \exp\!\left(
        -\frac{2k_0}{c_2}\,\tau
        + \frac{1}{8\sigma^2}
        + \frac{2\sigma^2 \tau^2}{c_2^2}
      \right).
\end{equation}
In particular, completing the square in the exponent, this can be written as
\begin{equation}
\langle a(\tau)\rangle
= \exp\!\left[
\frac{2\sigma^2}{c_2^2}\big(\tau-\tau_0\big)^2
-\frac{4k_0^2-1}{8\sigma^2}
\right],
\qquad 
\tau_0=\frac{k_0c_2}{2\sigma^2}.
\end{equation}
Hence $\langle a(\tau)\rangle$ exhibits a symmetric bounce centered at $\tau=\tau_0$, where the mean scale factor attains a strictly positive minimum,
\begin{equation}
a_{\min}=\langle a(\tau_0)\rangle
=\exp\!\left(-\frac{4k_0^2-1}{8\sigma^2}\right)>0,
\end{equation}
confirming that the quantum evolution is nonsingular at the level of expectation values. 
For large $|\tau|$, the quadratic term in the exponent dominates and one finds
\begin{equation}
\ln\langle a(\tau)\rangle \sim \frac{2\sigma^2}{c_2^2}\,\tau^2,
\qquad
\langle a(\tau)\rangle \propto 
\exp\!\left(\frac{2\sigma^2}{c_2^2}\tau^2\right).
\end{equation}
corresponding to a exponential expansion driven by the quantum spreading of the wave packet. 
Since $x=\ln a$ is Gaussian for all $\tau$, the scale factor $a=e^x$ follows a log-normal distribution. In particular all moments $\langle a^n\rangle$ are finite and the variance $\sigma_a^2=\langle a^2\rangle-\langle a\rangle^2$ remains finite for all~$\tau$, indicating that the quantum state stays well localized in minisuperspace throughout the evolution.

\section{Conclusion}

In this work we studied quantum cosmology in $f(R,T)$ gravity theory for an FLRW minisuperspace, with a particular emphasis on obtaining a time variable from the matter sector using Schutz perfect-fluid formalism. This choice is especially natural in $f(R,T)$ models, where the trace-dependent matter--geometry coupling makes matter an active component of the gravitational dynamics, and it allows the Wheeler--DeWitt framework to be recast into a Schr\"odinger--Wheeler--DeWitt (SWDW) evolution equation with respect to a matter clock $\tau$. 

We focused in particular on the minimally coupled sector $f(R,T)=F^0(R)+G^0(T)$, where the time construction used in earlier $f(R,T)$ quantizations based on the $(D,M)$-sector momentum is not available. In this setting, we derived the minisuperspace Hamiltonian and implemented Dirac quantization with a Hermitian factor-ordering prescription and a well-defined inner product, thereby making the probabilistic interpretation of the wave function explicit.

For the minimal sector $f(R,T)=F^0(R)+G^0(T)$, in the high-curvature regime $R\to\infty$ or equivalently in the early universe the coefficients entering the
SWDW equation are governed by the asymptotic ratio
\begin{equation}
\frac{B}{\mathcal{A}}
= \frac{G^{0}_{T}(T)}{\frac{1}{2} + 3\,G^{0}_{T}(T) + G^{0}_{TT}(T)\,\bigl(T - 4p\bigr)}\;\longrightarrow\;\varphi,
\end{equation}
which becomes a constant in the explicit examples considered, notably $\varphi=0$ and $\varphi=\tfrac{2}{7}$. We solved SWDW equations for representative choices of $F^0(R)$ and $G^0(T)$. Here are our remarks regarding the results:

\begin{itemize}
  \item
Setting $G^0(T)=0$ implies $B\equiv G^0_T(T)=0$ and hence $B/\mathcal{A}=0$, so the minimal $f(R,T)=F^0(R)+G^0(T)$ theory reduces to the usual $F^0(R)$ minisuperspace dynamics with Schutz time from the perfect fluid. In this limit the reduced Hamiltonian becomes
\begin{equation}
H \;=\; \frac{N}{6a^{3}}\bigl(AP_{A}^{2}-a\,P_{A}P_{a}\bigr)\;-\;6kNaA\;+\;Na^{3\alpha}P_{\tau}\;+\;Na^{3}V,
\qquad
V = F^0_{R}(R)\,R - F^0(R),
\end{equation}
in agreement with the standard $F^0(R)$ minisuperspace form with Schutz's fluid \cite{Vakali}. 
For the representative class $F^0(R)=b_1R+b_2R^n$ ($n\ge2$), introducing $x=a\sqrt{A}$ and $y=A$ leads to a reduced early-universe SWDW equation. For stiff matter, it becomes separable. In the flat case ($k=0$), the separated eigenfunctions are Bessel-type, and choosing seperation constant $\nu\ge0$ makes $\Psi\to0$ as $a\to0$, thus suppressing the classical singular boundary. Building normalizable wave packets via a quasi-Gaussian weight, one can compute expectation values in the many-worlds interpretation \cite{Everett}. In particular, for the illustrative $F^0(R)=b_2R^2$ model was obtained
\begin{equation}
\langle a(\tau)\rangle
=a_0\!\left[\gamma^2+\left(\frac{\tau}{12}\right)^2\right]^{\!\frac{1}{3(1+\alpha)}},
\qquad
a_{\min}=a_0\,\gamma^{\frac{1}{3(1+\alpha)}}>0,
\end{equation}
which exhibits a nonsingular bounce at $\tau=0$ and recovers the classical Friedmann scaling
$\langle a(\tau)\rangle\propto|\tau|^{\frac{2}{3(1+\alpha)}}$ for large $|\tau|$.
\item Setting {$F^0(R)=b_0+b_1R+b_2R^2$ and  $G^0(T)=\log|T|$}
implies $G_T^0(T)=1/T$. Consequently, we obtained $B=G^0_T(T)=1/T$. Along the early-universe trajectory $R\to\infty$ one also has
$|T|\to\infty$ (Appendix~A), hence
\begin{equation}
\varphi=\frac{B}{\mathcal{A}}\;\longrightarrow\;0,
\end{equation}
so the reduced SWDW dynamics again falls into the $\varphi=0$ class. In this limit the minisuperspace Hamiltonian
collapses to the standard $F^0(R)$ form with Schutz time,
\begin{equation}
H=\frac{N}{6a^{3}}\bigl(AP_{A}^{2}-aP_{A}P_{a}\bigr)-6kNaA+Na^{3\alpha}P_{\tau}+Na^{3}V,
\qquad
V=AR-f,
\end{equation}
therefore the logarithmic matter modification becomes dynamically negligible in the early universe. Since $G_T^0(T)=1/T$, the
minisuperspace dynamics is well-defined only on trajectories of fixed sign of $T$, and the logarithmic singularity at $T=0$
is dynamically avoided because $|T|\to\infty$ as $R\to\infty$. 

\item Setting {$G^0(T)=T$ implies $G_T^0(T)=1$ and $G_{TT}^0(T)=0$, hence the ratio
\begin{equation}
\varphi=\frac{B}{\mathcal{A}}=\frac{2}{7},
\end{equation}
is constant. Therefore, the reduced Hamiltonian acquires additional $\varphi$-dependent contributions.
Assuming that in the newborn quantum universe $C=F^0_{RR}(R)=0$, thus $F=\Box F^0_{RR}=0$, the theory becomes linear in $R$ and reduces to
$f(R,T)=b_1+b_2R+T$ with $A=b_2$, yielding
\begin{equation}
H= N \left[ \frac{7}{5} \left( -\frac{1}{b_2}\right) \frac{P_a^2}{24a}
- \frac{3}{7}a^{3}b_1- \frac{30}{7}\,kab_2
+ \frac{15}{7}\frac{P_{\tau}}{a^{3\alpha}}\right],
\end{equation}
and the corresponding SWDW equation on $(a,\tau)$ is separable via
$\Psi_{\mathcal{E}}(a,\tau)=e^{-i\mathcal{E}\tau}\psi(a)$.
In the flat stiff-fluid case $(k=0,\alpha=1)$, the reduced equation becomes Euler--Cauchy, and the Neumann condition selects the
vanishing branch at $a=0$,
\begin{equation}
\Psi_{\mathcal{E}}(a,\tau)= C_1\,e^{-i\mathcal{E}\tau}\,a^{\,1+\sqrt{\Delta}},
\end{equation}
therefore suppressing the classical singular boundary.
Direct wave-packet superposition in $a$ is non-normalizable with the minisuperspace measure, hence introducing
\begin{equation}
x=\ln a,\qquad \psi(a)=a\,\varphi(x),
\end{equation}
maps the stiff-fluid SWDW equation to a free-particle Schr\"odinger equation in $x$, allowing Gaussian spectral weights and a normalizable packet.
In this representation $x$ stays Gaussian for all $\tau$ and $a=e^{x}$ is log-normal. In particular,
\begin{equation}
\langle a(\tau)\rangle
=\exp\!\left[
\frac{2\sigma^2}{c_2^{\,2}}(\tau-\tau_0)^2-\frac{4k_0^2-1}{8\sigma^2}
\right],
\qquad 
\tau_0=\frac{k_0c_2}{2\sigma^2},
\qquad 
a_{\min}=\langle a(\tau_0)\rangle>0,
\end{equation}
which exhibits a quantum bounce at $\tau=\tau_0$.}
\end{itemize}

\newcommand{\bibTitle}[1]{``#1''}
\begingroup
\let\itshape\upshape
\bibliographystyle{plain}

\appendix
\section*{Appendix A}

Starting from the $f(R,T)$ field equations in units $16\pi=1$,
\[
f_{R}R_{\mu\nu}-\tfrac12 f\,g_{\mu\nu}+\bigl(g_{\mu\nu}\Box-\nabla_{\mu}\nabla_{\nu}\bigr)f_{R}
=\tfrac12 T_{\mu\nu}-f_{T}T_{\mu\nu}-f_{T}\Theta_{\mu\nu},
\]
their trace gives
\begin{equation}
f_{R}R-2f+3\Box f_{R}
=\tfrac12 T - f_{T}T - f_{T}\Theta.
\label{eq:trace_general}
\end{equation}
Then, one can obtain
\[
\Theta=+\,4p,
\]
and adopt the model
\[
f(R,T)=b_2R^{2}+b_1 R + b_0+\log|T|.
\]
With
\[
f_{R}=b_1+2b_2R,\qquad f_{T}=\frac{1}{T},
\]
Then, ~\eqref{eq:trace_general} becomes
\begin{align}
\bigl(b_1+2b_2R\bigr)R-2\bigl(b_2R^{2}+ b_1R + b_0+\log|T|\bigr)+3\Box(b_1+2b_2R)
&= \tfrac12 T - \frac{T}{T} - \frac{\Theta}{T} \nonumber\\
\Longrightarrow\quad
-\,b_1R-2b_0+6b_2\,\Box R - 2\log|T|
&= \tfrac12 T - 1 - \frac{4p}{T}.
\label{eq:master}
\end{align}
Equivalently,
\begin{equation}
-\,b_1R-2b_0+6b_2\,\Box R
=\tfrac12 T + 2\log|T| - 1 - \frac{4p}{T}.
\label{eq:master_alt}
\end{equation}
For a barotropic equation of state $p=\alpha\rho$ and perfect-fluid matter model, one has
\[
T=-\rho+3p=(-1+3\alpha)\rho
\quad\Rightarrow\quad
\frac{p}{T}=\frac{\alpha}{3\alpha-1}\qquad (\alpha\neq\tfrac13).
\]
Substituting into \eqref{eq:master_alt} yields
\begin{equation}
-\,b_1R-2b_0+6b_2\,\Box R
=\tfrac12 T + 2\log|T| - 1 - \frac{4\alpha}{3\alpha-1},
\end{equation}
or, multiplying by $2$ for convenience,
\begin{equation}
-\,2b_1R-4b_0+12b_2\,\Box R
= T + 4\log|T| - 2 - \frac{8\alpha}{3\alpha-1}.
\label{eq:final_balance}
\end{equation}
In FLRW cosmology, the dominant curvature term near the initial singularity is
\[
\Box R=\mathcal{O}\!\left(\frac{1}{a^{4}}\right)\;\longrightarrow\;\infty
\qquad (a\to 0),
\]
so, the left-hand side of \eqref{eq:final_balance} diverges with sign $\operatorname{sgn}(b\,\Box R)$ and magnitude
$\sim 12|b_2|\,|\Box R|$. To maintain \eqref{eq:final_balance}, the right-hand side must also diverge. 
Generically, this forces $|T|\to\infty$ as the linear $T$ term dominates the slow $\log|T|$ growth.
In the radiation limit $\alpha\to\tfrac13$ one has $T\to 0$, so $\log|T|\to -\infty$ and the right-hand side 
still blows up (in magnitude), keeping the balance with the divergent $\Box R$ term. Therefore, $B/\mathcal{A}\to 0$ when $R\to \infty$ or equivalently $|T| \to \infty$.

\end{document}